\documentstyle{mn}
\input{psfig}
\newif\ifAMStwofonts

\def\simgt{\hbox{\rlap{\raise 0.425ex\hbox{$>$}}\lower 0.65ex\h\label{Nbox{$\sim$}}}}
\def\simlt{\hbox{\rlap{\raise 0.425ex\hbox{$<$}}\lower 0.65ex\hbox{$\sim$}}}

\def\degree{^\circ}

\ifoldfss
  \ifCUPmtlplainloaded \else
    \NewTextAlphabet{textbfit} {cmbxti10} {}
    \NewTextAlphabet{textbfss} {cmssbx10} {}
    \NewMathAlphabet{mathbfit} {cmbxti10} {} 
    \NewMathAlphabet{mathbfss} {cmssbx10} {} 
  \fi
  \ifAMStwofonts
    \ifCUPmtlplainloaded \else
      \NewSymbolFont{upmath} {eurm10}
      \NewSymbolFont{AMSa} {msam10}
      \NewMathSymbol{\upi}     {0}{upmath}{19}
      \NewMathSymbol{\umu}     {0}{upmath}{16}
      \NewMathSymbol{\upartial}{0}{upmath}{40}
      \NewMathSymbol{\leqslant}{3}{AMSa}{36}
      \NewMathSymbol{\geqslant}{3}{AMSa}{3E}

      \let\leq=\leqslant 
       
    \fi
  \fi
\fi 

\ifnfssone
  \newmathalphabet{\mathit}
  \addtoversion{normal}{\mathit}{cmr}{m}{it}
  \addtoversion{bold}{\mathit}{cmr}{bx}{it}
  \newmathalphabet{\mathbfit} 
  \addtoversion{normal}{\mathbfit}{cmr}{bx}{it}
  \addtoversion{bold}{\mathbfit}{cmr}{bx}{it}
  \newmathalphabet{\mathbfss} 
  \addtoversion{normal}{\mathbfss}{cmss}{bx}{n}
  \addtoversion{bold}{\mathbfss}{cmss}{bx}{n}
  \ifAMStwofonts
    \ifCUPmtlplainloaded \else
      \UseAMStwoboldmath
      \makeatletter
      \new@mathgroup\upmath@group
      \define@mathgroup\mv@normal\upmath@group{eur}{m}{n}
      \define@mathgroup\mv@bold\upmath@group{eur}{b}{n}
      \edef\UPM{\hexnumber\upmath@group}
      \new@mathgroup\amsa@group
      \define@mathgroup\mv@normal\amsa@group{msa}{m}{n}
      \define@mathgroup\mv@bold\amsa@group{msa}{m}{n}
      \edef\AMSa{\hexnumber\amsa@group}
      \makeatother
      \mathchardef\upi="0\UPM19
      \mathchardef\umu="0\UPM16
      \mathchardef\upartial="0\UPM40
      \mathchardef\leqslant="3\AMSa36
      \mathchardef\geqslant="3\AMSa3E

      \let\leq=\leqslant 

    \fi
  \fi
\fi 

\ifnfsstwo
  \DeclareMathAlphabet{\mathbfit}{OT1}{cmr}{bx}{it}
  \SetMathAlphabet\mathbfit{bold}{OT1}{cmr}{bx}{it}
  \DeclareMathAlphabet{\mathbfss}{OT1}{cmss}{bx}{n}
  \SetMathAlphabet\mathbfss{bold}{OT1}{cmss}{bx}{n}
  \ifAMStwofonts
    \ifCUPmtlplainloaded \else
      \DeclareSymbolFont{UPM}{U}{eur}{m}{n}
      \SetSymbolFont{UPM}{bold}{U}{eur}{b}{n}
      \DeclareSymbolFont{AMSa}{U}{msa}{m}{n}
      \DeclareMathSymbol{\upi}{0}{UPM}{"19}
      \DeclareMathSymbol{\umu}{0}{UPM}{"16}
      \DeclareMathSymbol{\upartial}{0}{UPM}{"40}
      \DeclareMathSymbol{\leqslant}{3}{AMSa}{"36}
      \DeclareMathSymbol{\geqslant}{3}{AMSa}{"3E}

      \let\leq=\leqslant 

    \fi
  \fi
\fi 

\ifCUPmtlplainloaded \else
  \ifAMStwofonts \else 
    \def\upi{\pi}
    \def\umu{\mu}
    \def\upartial{\partial}
  \fi
\fi

\def \Om {\Omega_0}
\def \lo {\lambda_0}
\def \bj {b_{\rm J}}
\def \kms {{\rm km\,s}$^{-1}$}
\def \lam {$\lambda$}
\def \ew {W_{\lambda}}
\def \ewg {W_{\lambda {\rm gal}}}
\def \lgal {L_{\rm gal}}
\def \lqso {L_{\rm QSO}}
\def \ltot {L_{\rm Tot}}
\def \mgal {M_{\rm gal}}
\def \mqso {M_{\rm QSO}}
\def \mtot {M_{\rm Tot}}

\def \mb {M_{\rm B}}
\def \la {Ly$\alpha$}
\def \ha {H$\alpha$}
\def \hb {H$\beta$}
\def \hg {H$\gamma$}
\def \hd {H$\delta$}
\def \hep {H$\epsilon$}
\def \oiii {[O{\small~III}]}
\def \oii {[O{\small~II}]}

\def \neiii {[Ne{\small~III}]}
\def \nev {[Ne{\small~V}]}
\def \mgii {Mg{\small~II}}
\def \caii {Ca{\small~II}}
\def \cak {Ca{\small~II}~K}
\def \cah {Ca{\small~II}~H}
\def \civ {C{\small~IV}}
\def \ciii {C{\small~III}]}
\def \siiv {Si{\small~IV}}
\def \siiii {Si{\small~III}]}
\def \oiv {O{\small~IV}}
\def \feii {Fe{\small~II}}
\def \nv {N{\small~V}}
\def \hei {He{\small~I}}
\def \heii {He{\small~II}}
\def \aliii {Al{\small~III}}

\def \centrho {\multicolumn{1}{c}{$\rho$}}
\def \centp {\multicolumn{1}{c}{$P$}}
\def \aj {AJ}
\def \mnras {MNRAS}
\def \apj {ApJ}

\def \aap {A\&A}

\title[QSO line strength vs. luminosity and redshift]{The correlation
of line strength with luminosity and redshift from composite QSO spectra} 
\author[S. M. Croom et al.]
{S. M. Croom$^1$\thanks{scroom@aaoepp.aao.gov.au}, K. Rhook$^1$,
  E. A. Corbett$^1$, B. J. Boyle$^1$, H. Netzer$^2$, N. S. Loaring$^3$,
\newauthor L. Miller$^3$, P. J. Outram$^4$, T. Shanks$^4$, R. J. Smith$^5$ \\
$^1$ Anglo-Australian Observatory, PO Box 296, Epping, NSW 1710, Australia\\
$^2$ School of Physics and Astronomy, Tel-Aviv University, Tel-Aviv
69978, Israel\\
$^3$Department of Physics, Oxford University, Keble Road, Oxford, OX1
3RH, UK\\
$^4$Physics Department, University of Durham, South Road, Durham, DH1 3LE,
UK\\
$^5$Astrophysics Research Institute, Liverpool John Moores University,
Twelve Quays House, Egerton Wharf, Birkenhead,  CH41 1lD, UK}

\begin{document}

\maketitle

\begin{abstract}
We have generated a series of composite QSO spectra using over 22000
individual low resolution ($\sim8$\AA) QSO spectra obtained from the
2dF ($18.25<\bj<20.85$) and 6dF ($16<\bj\leq18.25$) QSO Redshift
Surveys. The large size of the catalogue has enabled us to construct
composite spectra in relatively narrow redshift ($\Delta z=0.25$) and
absolute magnitude ($\Delta \mb=0.5$) bins.  The median number of QSOs
in each composite spectra is $\sim200$, yielding typical
signal-to-noise ratios of $\sim100$. For a given redshift interval, the
composite spectra cover a factor of over 25 in luminosity.  For a
given luminosity, many of the major QSO emission lines (e.g. \mgii\
\lam2798, \oii\ \lam3727) can be observed over a redshift range of one
or greater.  

Using the composite spectra we have measured the line strengths
(equivalent widths) of the major broad and narrow emission lines.  We
have also measured the equivalent width of the \caii\ \lam3933 K
absorption feature due to the host galaxy of the AGN.  Under the
assumption of a fixed host galaxy spectral energy distribution (SED),
the correlation seen between \cak\ equivalent width and source
luminosity implies $\lgal\propto\lqso^{0.42\pm0.05}$.  We find strong
anti-correlations with luminosity for the equivalent widths of \oii\
\lam3727 and \nev\ \lam3426. These provide hints to the general fading
of the NLR in high luminosity sources which we attribute to the NLR
dimensions becoming larger than the host galaxy. This could have
important implications for the search for type 2 AGN at high
redshifts.  If average AGN host galaxies have SEDs similar to average
galaxies, then the observed narrow \oii\ emission could be solely due
to the host galaxy at low luminosities ($\mb\sim-20$).  This suggests
that the \oii\ line observed in high luminosity AGN may be emitted, in
a large part, by intense star-forming regions. The AGN contribution to
this line could be weaker than previously assumed.

We measure highly significant Baldwin effects for most broad emission
lines (\civ\ \lam1549, \ciii\ \lam1909, \mgii\ \lam2798, \hg, \hb) and
show that they are predominantly due to correlations with luminosity,
not redshift.  We find that the \hb\ and \hg\ Balmer lines show an
{\it inverse Baldwin effect} and are positively correlated with
luminosity, unlike the broad UV lines.  We postulate that this
previously unknown effect is due to a luminosity dependent change in
the the ratio of disk to non-disk continuum components.
\end{abstract}


\section{introduction}

The correlation of QSO emission line properties with luminosity is a
straightforward yet potentially highly informative test of standard
physical models for AGN.  Since the discovery of an anti-correlation
between the equivalent width ($\ew$) of the \civ\ \lam1549 emission
line and continuum luminosity ($L$) by Baldwin (1977)\nocite{b77}, a
significant amount of effort  has been expended quantifying this
relationship (hereinafter referred to as the Baldwin effect), and
investigating similar correlations with other QSO emission lines
(Baldwin, Wampler \& Gaskell 1989; Zamorani et al.\ 1992; Green,
Forster \& Kuraszkiewcz 2001)\nocite{bwg89,zam92,gfk01}.  The results
have revealed that the anti-correlation with luminosity is relatively
weak, typically $\ew\propto L^{\beta}$, with $\beta=-0.2$ and large
scatter.  Similar correlations have been seen in most other broad
emission lines including \mgii\ \lam2798, \ciii\ \lam1909, \siiv+\oiv]
\lam1400 and \la\ with $-0.4<\beta<-0.1$ \cite{gfk01}.  It has also
been claimed \cite{gfk01} that the Baldwin  effect may be dominated by
an even stronger anti-correlation with redshift.  However, in the
magnitude-limited QSO samples that have been studied to date, it is
extremely difficult to disentangle the effects of redshift and
luminosity.  It it typically only possible to access 1--1.5 magnitudes
at any given redshift, given the steep slope of the QSO luminosity
function for magnitude-limited samples with $B<19.5$ (Boyle, Shanks \&
Peterson 1988).

A further limitation of existing studies is that it is difficult to
study the correlation with luminosity and/or redshift for weaker
lines, in particular the narrow line region (NLR).  The spectra used
in such analyses are typically 'survey' quality, i.e. relatively low
signal-to-noise ratio ($SNR\sim5-10$) and thus narrow emission lines
can be difficult to detect in individual spectra.  

Composite QSO spectra have been generated from most large QSO surveys
over the past decade \cite{b90,fran91}, providing a detailed picture
of the ensemble average spectral properties of the QSO sample.
Typical $SNR$s in these spectra approach or even exceed 100, with even
relatively weak emission lines (e.g. \nev\ \lam3426) easily
detectable.  However, previous surveys have been too small (comprising
1000 QSOs or less) to generate composite spectra as a function of both
luminosity and redshift with which to examine correlations.  With the
recent advent of much larger QSO surveys such as the 2dF QSO Redshift
Survey (2QZ, Croom et al.\ 2001)\nocite{2qzpaper5} and and Sloan
Digital Sky Survey (SDSS, Vanden Berk et al.\ 2001; Schneider et al.\
2002)\nocite{sdssqso,sdsscomp} we may now use  composite, rather than
individual spectra, to investigate the correlation of QSO spectral
properties with luminosity and redshift in much greater detail that
has hitherto been possible.

In this paper we describe the result of an analysis of composite QSO
spectra based on the almost 22000 QSOs observed to date (January 2002)
in the 2QZ.  The bulk of these objects lie around the break in the
luminosity function, thus providing a better sampling in luminosity at
any given redshift than QSO surveys at the bright end of the LF
(e.g. the Large Bright Quasar Survey; Hewett, Foltz \& Chaffee 1995).
Moreover, we have also included a few hundred brighter  QSOs observed
with the new 6-degree Field (6dF) multi-object spectrographic facility
on the UK Schmidt Telescope (Croom et al. in preparation) to increase
the luminosity range studied at any given redshift to typically 3--4
magnitudes.  As well as providing a wide baseline over which to study
correlations such as the Baldwin effect, this sampling of the QSO
$(L,z)$ plane provides an opportunity to disentangle the effects of
luminosity and redshift.

In Section 2 we describe the data used in our analysis, while in
Section 3 we discuss the methods used to generate the
composite spectra and measure the spectral line equivalent
widths.  In section 4 we present the results of our analysis, we then
discuss these in the context of theoretical models in section 5. 

\section{Data}

\subsection{Generation of composite spectra}

\begin{table*} 
\caption{The number of QSOs in each of our absolute magnitude-redshift
($\mb-z$) bins.  For each bin the central redshift and absolute magnitude
is displayed.  The last column shows the total number of QSOs
in each magnitude interval over all redshifts.  In some $\mb-z$
intervals there are only a small number of QSOs.  In these cases the
spectra in adjacent $\mb$ intervals were combined together, an $^*$ or
$^\dagger$ indicates where this has been done.  For example, in the
$z=0.375$ interval, QSOs in the $\mb=-24.75,-24.25$ and --23.75 bins
were combined together.}
\label{table_bins}
\begin{center}
\begin{tabular}{rrrrrrrrrrrrrr}
\hline
&\multicolumn{13}{c}{Redshift}\\
$\mb$   &0.125&0.375&0.625&0.875&1.125&1.375&1.625&1.875&2.125&2.375&2.625&2.875& all \\
\hline										        
--29.25 &  -- &  -- &  -- &  -- &  -- &  -- &  -- &  -- &  -- &   1$^*$ &  -- &  -- &   1$^*$ \\
--28.75 &  -- &  -- &  -- &  -- &  -- &  -- &  -- &  -- &  -- &  -- &   1$^*$ &  -- &   1$^*$  \\
--28.25 &  -- &  -- &  -- &  -- &  -- &  -- &  -- &  -- &   1$^*$ &   1$^*$ &   2$^*$ &   3$^*$ &   7$^*$  \\
--27.75 &  -- &  -- &  -- &  -- &  -- &  -- &   2$^*$ &   1$^*$ &   8$^*$ &  13$^*$ &  15$^*$ &   4$^*$ &  43$^*$  \\
--27.25 &  -- &  -- &  -- &  -- &   1$^*$ &   3$^*$ &   5$^*$ &  23$^*$ &  48$^*$ &  48 &  37$^*$ &  13$^*$ & 180 \\
--26.75 &  -- &  -- &  -- &   1$^*$ &   5$^*$ &  10$^*$ &  54 & 100 & 152 & 116 & 101 &  37$^*$ & 585 \\
--26.25 &  -- &  -- &  -- &   3$^*$ &  13$^*$ &  80 & 183 & 259 & 316 & 289 & 211 &  71 &1432 \\
--25.75 &  -- &  -- &   3$^*$ &  12$^*$ &  76 & 194 & 370 & 517 & 545 & 439 & 277 &  50 &2483 \\
--25.25 &  -- &  -- &   4$^*$ &  77 & 218 & 434 & 667 & 788 & 791 & 446 &  99 &  -- &3524 \\
--24.75 &  -- &   2$^*$ &  27$^*$ & 178 & 441 & 766 & 956 & 942 & 341 &  25 &  -- &  -- &3678 \\
--24.25 &  -- &   7$^*$ &  85 & 360 & 703 & 974 & 779 &  92 &  -- &  -- &  -- &  -- &3000 \\
--23.75 &  -- &  14$^*$ & 217 & 546 & 778 & 456 &  11 &  -- &  -- &  -- &  -- &  -- &2022 \\
--23.25 &   2$^*$ &  66 & 334 & 692 & 359 &  -- &  -- &  -- &  -- &  -- &  -- &  -- &1453 \\
--22.75 &  -- & 101 & 410 & 420 &   2 &  -- &  -- &  -- &  -- &  -- &  -- &  -- & 933 \\
--22.25 &   2$^*$ & 158 & 528 &  51 &  -- &  -- &  -- &  -- &  -- &  -- &  -- &  -- & 739 \\
--21.75 &   8$^*$ & 220 & 274 &  -- &  -- &  -- &  -- &  -- &  -- &  -- &  -- &  -- & 502 \\
--21.25 &  15$^\dagger$ & 239 &  38 &  -- &  -- &  -- &  -- &  -- &  -- &  -- &  -- &  -- & 292 \\
--20.75 &  22$^\dagger$ & 117 &  -- &  -- &  -- &  -- &  -- &  -- &  -- &  -- &  -- &  -- & 139 \\
--20.25 &  22$^\dagger$ &  27$^*$ &  -- &  -- &  -- &  -- &  -- &  -- &  -- &  -- &  -- &  -- &  49 \\
--19.75 &  16$^*$ &   7$^*$ &  -- &  -- &  -- &  -- &  -- &  -- &  -- &  -- &  -- &  -- &  23$^*$  \\
--19.25 &  11$^*$ &  -- &  -- &  -- &  -- &  -- &  -- &  -- &  -- &  -- &  -- &  -- &  11$^*$  \\
--18.75 &   4$^*$ &  -- &  -- &  -- &  -- &  -- &  -- &  -- &  -- &  -- &  -- &  -- &   4$^*$  \\
--18.25 &  -- &  -- &  -- &  -- &  -- &  -- &  -- &  -- &  -- &  -- &  -- &  -- &  -- \\
--17.75 &   1$^*$ &  -- &  -- &  -- &  -- &  -- &  -- &  -- &  -- &  -- &  -- &  -- &   1$^*$  \\
\hline
\end{tabular}
\end{center}
\end{table*}

The data used in our analysis is taken from the 2dF and 6dF QSO
Redshift Surveys (2QZ, Croom et al 2001; 6QZ Croom et al. in
preparation).  QSO candidates were selected for observation based on
their stellar appearance and blue colours found from APM measurements
of UK Schmidt Telescope (UKST) photographic plates and films in the
$u$, $\bj$ and $r$ bands.  The 2QZ/6QZ area comprises 30 UKST fields
arranged in two $75\degree\times5\degree$ declination strips centred
on $\delta=-30\degree$ and $\delta=0\degree$.  The $\delta=-30\degree$
strip extends from $\alpha=21^h40$ to $\alpha=3^h15$ in the South
Galactic Cap and the equatorial strip from $\alpha=9^h50$ to
$\alpha=14^h50$ in the North Galactic Cap.  The 2QZ and 6QZ sources
were selected from the same photometric data, the only difference
being their ranges in apparent magnitude: $18.25<\bj<20.85$ (2QZ)
and $16.0<\bj\leq18.25$ (6QZ).  The combined datasets thus produce a
uniform QSO sample over a wide range in luminosity.  Details of the
candidate selection can be found in Smith et al. (2002).

The 2QZ objects were observed over the period October 1997 to January
2002 using the 2dF instrument at the Anglo-Australian Telescope.
Observations were made with the low dispersion 300B grating, providing
a dispersion of 178.8~\AA~mm$^{-1}$ (4.3~\AA~pixel$^{-1}$) and a
resolution of $\simeq8.6$~\AA\ over the range 3700--7900\AA.  Typical
integration times were 55 minutes, in a range of observing conditions
(1--2.5 arcsec seeing) resulting in median $SNR\sim5$ per pixel.  The
brighter 6QZ objects used in the present paper were observed in
September 2001 using the 6dF facility at the UKST.  A low dispersion
250B grating was used to provide a dispersion of 286~\AA~mm$^{-1}$
(3.6~\AA~pixel$^{-1}$) and a resolution of $\simeq11.3$~\AA\ over the
range 3900--7600\AA.  Exposure times were typically 100
minutes resulting in median $SNR\sim15$ per pixel.

Data from both 2dF and 6dF were reduced using the pipeline data
reduction system {\small 2DFDR} \cite{bailey02}.  Identification of
spectra and the determination of redshifts was carried out by a
automated program, {\small AUTOZ} (Croom et al. 2001; Miller et al. in
preparation).  Each spectrum was checked by eye by two members of the
team.  In our analysis below we only include quality class 1
identifications (96 per cent reliable identification), these being the
best quality spectra.  We also only take the best spectrum (based on
quality class and then $SNR$) of each object in the case where there
is more than one spectrum available.  The combined 2QZ/6QZ dataset
provides us with 22041 independent QSO spectra.

Typical redshift errors are $\sigma_{\rm z}=0.003$ and photometric
errors in the $\bj$ band are $\sim0.1$ mag.  Absolute magnitudes were
computed from the observed photographic $\bj$ magnitude, after
correction for Galactic extinction \cite{sfd98}, using the
K-corrections found by Cristiani \& Vio (1990)\nocite{cv90}.
Throughout we assume a flat cosmological world model with $\Om=0.3$, 
$\lo=0.7$ and $H_0=70$~\kms~Mpc$^{-1}$.

\begin{table} 
\caption{List of strong spectral features removed before continuum
fitting.  A simple linear interpolation is made between two
'continuum' bands defined on either side of the feature.}
\label{table_cont}
\begin{center}
\begin{tabular}{lll}
\hline
            & blue cont. & red cont. \\
Feature     & band (\AA) & band (\AA)\\
\hline
\la+\nv     & 1130--1155 & 1280--1290\\
\siiv+\oiv] & 1350--1360 & 1445--1470\\
\civ+\heii  & 1445--1470 & 1685--1705\\
\ciii+\aliii& 1800--1830 & 1985--2020\\
\mgii+\feii & 2650--2685 & 3025--3065\\
\oii        & 3675--3705 & 3745--3785\\
\neiii      & 3845--3855 & 3905--3920\\
\hd         & 4020--4050 & 4165--4200\\
\hg         & 4220--4270 & 4430--4460\\
\feii       & 4430--4460 & 4710--4760\\
\hb+\oiii   & 4710--4760 & 5080--5105\\
\feii       & 5080--5105 & 5450--5500\\
\hei        & 5740--5790 & 5940--5980\\
\ha         & 6320--6380 & 6745--6805\\
\hline
\end{tabular}
\end{center}
\end{table}

\section{Method}

We have generated composite QSO spectra in discrete absolute magnitude
($\Delta M_B=0.5\,$mag) and redshift ($\Delta z = 0.25$) bins.  The bin
widths were chosen to give good resolution in luminosity and redshift,
whilst typically retaining over 100 QSOs in at least 5 magnitude bins
(a factor of 10 in luminosity) at each redshift (see Table
\ref{table_bins}).  Once QSOs identified as broad absorption line (BAL)
QSOs had been removed, there remained a total of 21102
QSOs with which to generate the composite spectra.  The most important
issue relating to the construction of the composites was that the
spectra were not flux calibrated.  The effects of differential
atmospheric refraction, corrector chromatic aberration and fibre
positioning errors makes obtaining even a relative flux calibration
for sources extremely challenging.  We therefore chose not to attempt
flux calibration of our spectra.  We did however correct for
absorption due to the atmospheric telluric bands (the optical fibres
also provide some absorption in these same bands).  We summed all the
spectra in a single observation in order to obtain a mean absorption
correction which was then applied to the data.  Also, pixels which had
anomalously high variance due to residuals of night sky emission lines
were flagged as bad and discarded from our analysis.

As the spectra were not flux calibrated we decided to normalize each
spectrum to a continuum level as a function of wavelength.  This
allows us to measure equivalent widths, line widths and line centres,
however we lose any information concerning continuum shape and
absolute line strengths.  Fitting the continuum relies on defining
line-free parts of the spectrum.  This is not always possible,
particularly in regions of the spectrum dominated by weak \feii\
emission.  Our approach, therefore, was to remove all strong emission
line features, interpolating linearly between pseudo-continuum bands
defined on each side of the line.  The strong features
removed, and the continuum bands defined are listed in Table
\ref{table_cont}.  After removing these strong lines, a 4th order
polynomial was fit to each spectrum, which was then used to divide
the spectra, providing an approximate continuum
normalization.  In a second step to remove residual large-scale
features in the spectrum, each spectrum was divided by a median
filtered version using a wide box-car filter of width 201 pixels (each
spectrum containing 1024 pixels or 1032 pixels for 2dF and 6dF data
respectively).  At the edges of the spectrum the filter was reduced in
size to a minimum half-width of 5 pixels.

\begin{figure*}
\centering
\centerline{\psfig{file=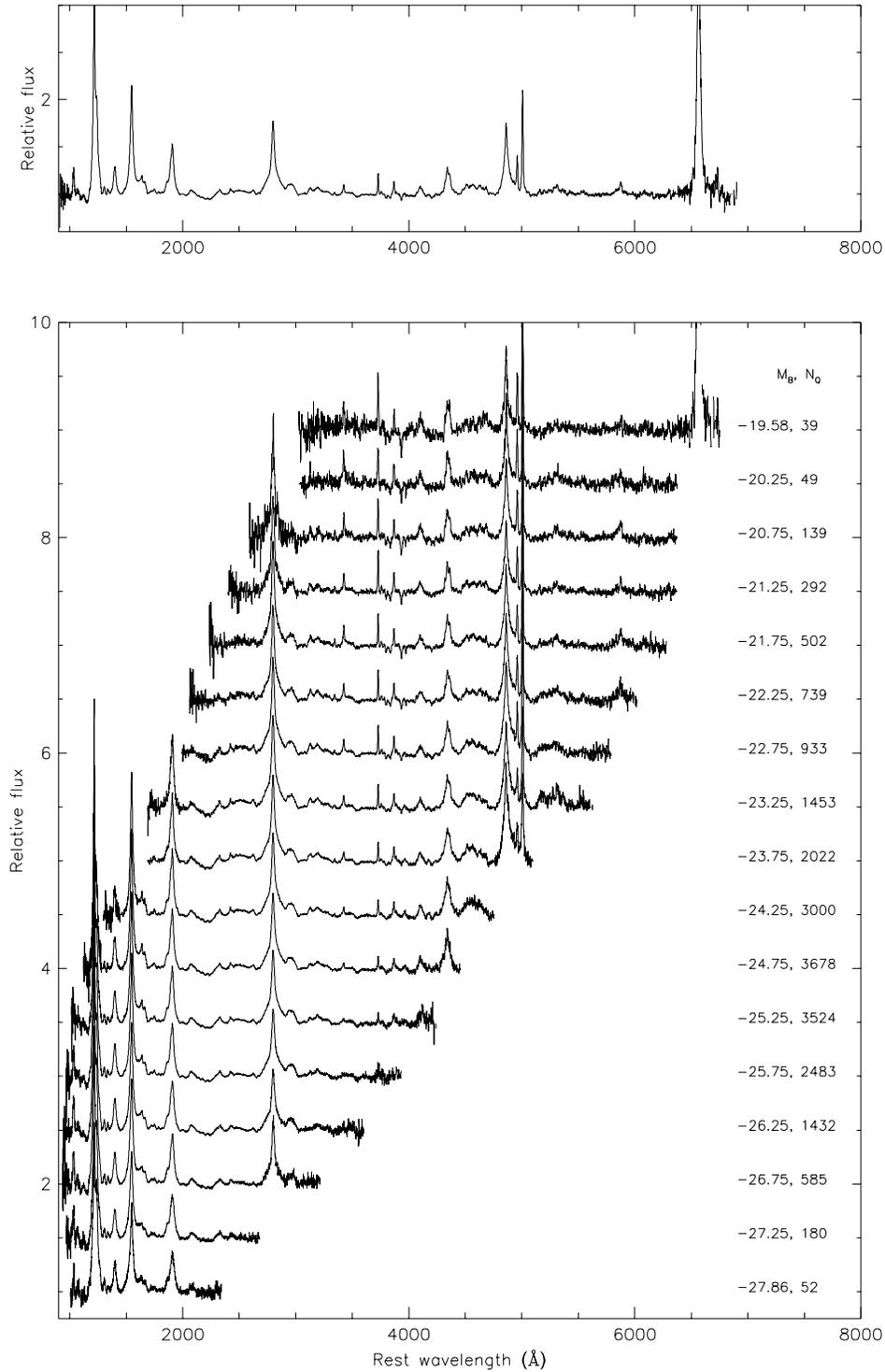,width=16cm}}
\caption{Composite QSO spectra.  Top: all spectra combined into one
composite.  Bottom: composites computed in absolute magnitude
intervals ($\Delta\mb=0.5$) with no redshift binning.  The brightest
and faintest bins have been made wider to include a sufficient number
of spectra.  All the spectra have a continuum level of one, but have
been offset for clarity.}
\label{fig_lumcomp}
\end{figure*}

\begin{figure*}
\centering
\centerline{\psfig{file=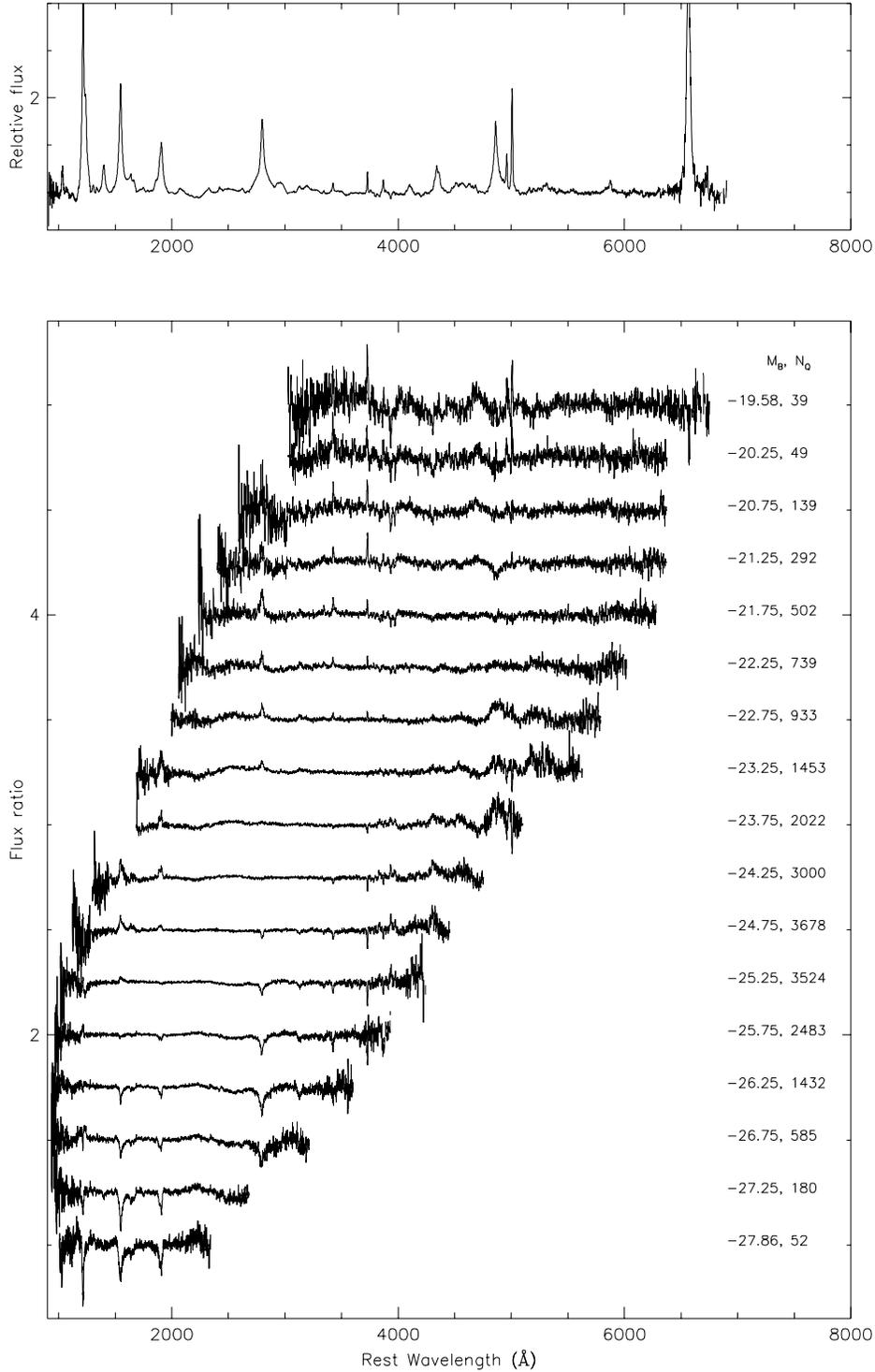,width=16cm}}
\caption{The luminosity segregated composite spectra, as shown in
Fig. \ref{fig_lumcomp}, divided by the average composite (top).  From
this apparent correlations can be seen between luminosity and line
strength in a number of lines, including \oii, \nev, \mgii, \ciii\ and
\civ.  These correlations are discussed in the text.  Again, the mean
flux ratio in each spectrum is one, but the spectra have been offset
for clarity.}
\label{fig_ratio}
\end{figure*}

\begin{figure*}
\centering
\centerline{\psfig{file=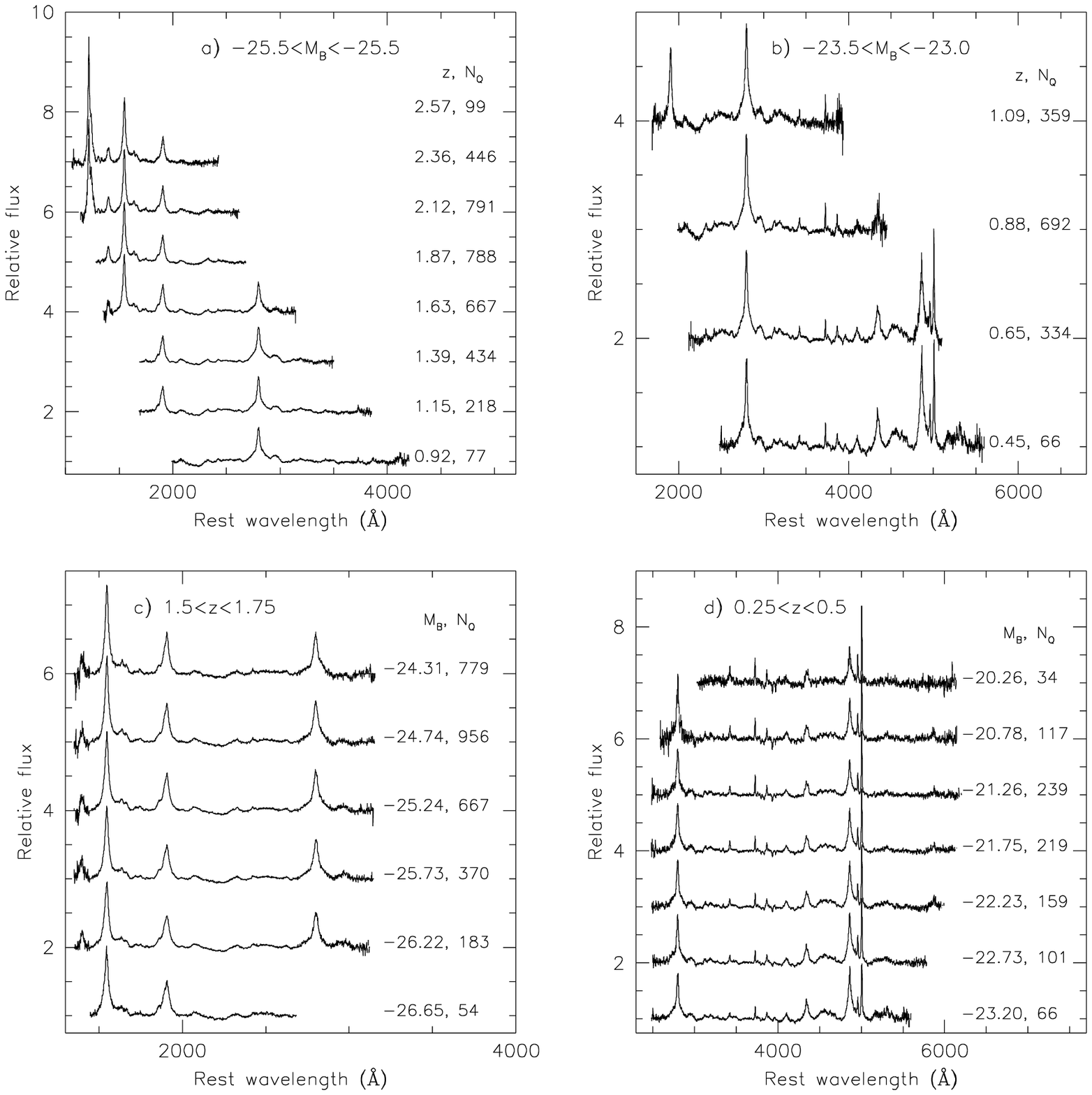,width=16cm}}
\caption{Examples of the QSO composites generated in absolute
magnitude-redshift intervals.  Top: constant luminosity examples
($-25.5<\mb<-25.0$ and $-23.5<\mb<-23.0$) over a range of redshifts.
Bottom: constant redshift intervals as a function of luminosity for
$1.5<z<1.75$ and $0.25<z<0.5$.} 
\label{fig_lzcomp}
\end{figure*}

The above processing was all carried out in the observed frame.  After
continuum normalization the spectra were shifted to the rest frame,
interpolating linearly onto a uniform scale of 1\AA\,${\rm
pixel}^{-1}$.  Finally the composite spectra were produced by taking
the median value of each pixel.  For each pixel the median $z$ and
$\mb$ of the contributing QSOs was also determined.  We can then
determine apropriate values for each feature, and not just each
composite.  The values of $z$ and $\mb$ assigned to each feature are
the average, over the wavelength range of the feature, of the pixel
median $z$ and $\mb$ values.  We derived errors for each composite by
looking at the distribution of values to be medianed for each pixel.
The 1$\sigma$ errors were taken to be the 68\% semi-interquartile
range of the pixel values divided by the square root of the number of
objects contributing.  We have constructed composites in $z-\mb$ and
also composites binned in absolute magnitude only.  One final
composite was made from all the spectra (see
Fig. \ref{fig_lumcomp}). The resulting composites are normalized to
the pseudo-continuum over most parts of the spectrum, except for the
2000--3500\AA\ region where our procedure treats the \feii\ emission
bands as though they were continuum.  Therefore, in our subsequent
analysis of these spectra we are unable to deduce any results
concerning these broad \feii\ features.

\begin{table*} 
\caption{Spectral features studied. The first column gives the
principal emission line in each feature for which line equivalent
widths were measured. Columns 2 and 3 give the regions of the spectrum
used in the continuum fit.  Columns 4 to 10 give the properties of
the individual fitted components.  Column 4 lists the component
number, column 5 the element/elements causing the emission, column 6
the laboratory (vacuum) wavelengths of the components and column 7
indicates whether the emission is narrow or broad.  Columns 8, 9 and 10
list the fitted parameters, showing which parameters were tied
together.  Notes: 1. Unresolved doublets or multiplets, mean
wavelength quoted.  2. Although this feature is identified
here as \heii\ it is actually a blend of several lines including
\feii\ and OIII] and is therefore relatively broad. 3. The \neiii\
\lam3968 feature is also contaminated by the \hep\ \lam3970 which may
be present in emission or absorption. 4. When both \hg\ and \hb\ were
present in the spectrum, the velocity width of these components were
fixed to that measured for the \oiii\ \lam5007 emission line.}
\label{table_fits}
\begin{center}
\begin{tabular}{lccccccccc}
\hline
Principal  &Blue Cont.  & Red Cont.  & Component & Emission  & $\lambda_{\rm lab}$& emission &$\lambda_{\rm c}$ & amp. & $\sigma$\\  
Line       & \AA        & \AA        & number    & source    & \AA                & type     &                  &      &         \\  
\hline
\la        & 1135--1155 & 1320--1340 & 1 & \la      & 1215.67 & Broad  & $\lambda_1$ & $a_1$ & $\sigma_1$ \\
           &            &            & 2 & \la      & 1215.67 & Broad  & $\lambda_2$ & $a_2$ & $\sigma_2$ \\
           &            &            & 3 & \nv      & 1240.14 & Broad  & $\lambda_1$ & $a_3$ & $\sigma_1$ \\
           &            &            & 4 & \nv      & 1240.14 & Broad  & $\lambda_1$ & $a_4$ & $\sigma_2$ \\
\hline
\siiv+\oiv & 1350--1365 & 1440--1455 & 1 & \siiv$^1$& 1396.76 & Narrow & $\lambda_1$ & $a_1$ & $\sigma_1$ \\ 
           &            &            & 2 & \siiv$^1$& 1396.76 & Broad  & $\lambda_2$ & $a_2$ & $\sigma_2$ \\
           &            &            & 3 & \siiv$^1$& 1396.76 & Broad  & $\lambda_2$ & $a_3$ & $\sigma_3$ \\
\hline
\civ       & 1440--1460 & 1690--1710 & 1 & \civ$^1$ & 1549.06 & Narrow & $\lambda_1$ & $a_1$ & $\sigma_1$ \\
           &            &            & 2 & \civ$^1$ & 1549.06 & Broad  & $\lambda_2$ & $a_2$ & $\sigma_2$ \\
           &            &            & 3 & \civ$^1$ & 1549.06 & Broad  & $\lambda_2$ & $a_3$ & $\sigma_3$ \\
           &            &            & 4 & \heii$^2$& 1640.42 & Broad  & $\lambda_4$ & $a_4$ & $\sigma_4$ \\
\hline
\ciii      & 1800--1820 & 1975--1995 & 1 & \ciii    & 1908.73 & Narrow & $\lambda_1$ & $a_1$ & $\sigma_1$ \\
           &            &            & 2 & \ciii    & 1908.73 & Broad  & $\lambda_2$ & $a_2$ & $\sigma_2$ \\
           &            &            & 3 & \ciii    & 1908.73 & Broad  & $\lambda_2$ & $a_3$ & $\sigma_3$ \\ 
           &            &            & 4 & \aliii   & 1857.40 & Broad  & $\lambda_2$ & $a_4$ & $\sigma_2$ \\ 
           &            &            & 5 & \siiii   & 1892.03 & Broad  & $\lambda_2$ & $a_5$ & $\sigma_2$ \\
\hline
\mgii      & 2640--2660 & 3030--3050 & 1 & \mgii$^1$& 2798.75 & Broad  & $\lambda_1$ & $a_1$ & $\sigma_1$ \\
           &            &            & 2 & \mgii$^1$& 2798.75 & Broad  & $\lambda_2$ & $a_2$ & $\sigma_2$ \\
           &            &            & 3 &\feii\ blend&2965   & Broad  & $\lambda_3$ & $a_3$ & $\sigma_3$ \\
\hline
\nev       & 3360--3380 & 3450--3470 & 1 & \nev     & 3426.84  & Narrow & $\lambda_1$ & $a_1$ & $\sigma_1$ \\
           &            &            & 2 & \feii?   & 3415    & Narrow & $\lambda_1$ & $a_1$ & $\sigma_2$ \\
\hline
\oii       & 3700--3710 & 3742--3752 & 1 & \oii$^1$ & 3728.48 & Narrow & $\lambda_1$ & $a_1$ & $\sigma_1$ \\
\hline
\neiii     & 3845--3850 & 3910--3915 & 1 & \neiii   & 3869.85 & Narrow & $\lambda_1$ & $a_1$ & $\sigma_1$ \\
           &            &            & 2 & \hei     & 3889.74 & Broad  & $\lambda_2$ & $a_2$ & $\sigma_2$ \\ 
\hline
\cak       & 3900--3910 & 4010--4020 & 1 & \cak     & 3934.78 & Broad  & $\lambda_1$ & $a_1$ & $\sigma_1$ \\
           &            &            & 2 & \cah     & 3969.59 & Broad  & $\lambda_1$ & $a_1$ & $\sigma_1$ \\
           &            &            & 3 &\neiii$^3$& 3968.58 & Narrow & $\lambda_3$ & $a_3$ & $\sigma_3$ \\
\hline
\hd        & 4000--4020 & 4200--4220 & 1 & \hd      & 4102.89 & Narrow & $\lambda_1$ & $a_1$ & $\sigma_1$ \\
           &            &            & 2 & \hd      & 4102.89 & Broad  & $\lambda_2$ & $a_2$ & $\sigma_2$ \\
\hline
\hg        & 4200--4220 & 4440--4460 & 1 & \hg      & 4341.68 & Narrow & $\lambda_1$ & $a_1$ & $\sigma_1^4$ \\
           &            &            & 2 & \hg      & 4341.68 & Broad  & $\lambda_2$ & $a_2$ & $\sigma_2$ \\
           &            &            & 3 & \oiii+Fe & 4361.62 & Narrow & $\lambda_2$ & $a_3$ & $\sigma_1^4$ \\ 
\hline
\hb        & 4740--4760 & 5070--5090 & 1 & \hb      & 4862.68 & Narrow & $\lambda_1$ & $a_1$ & $\sigma_1$ \\
           &            &            & 2 & \hb      & 4862.68 & Broad  & $\lambda_2$ & $a_2$ & $\sigma_2$ \\
           &            &            & 3 & \oiii    & 4960.30 & Narrow & $\lambda_1$ & $a_3$ & $\sigma_1$ \\
           &            &            & 4 & \oiii    & 5008.24 & Narrow & $\lambda_1$ & $a_4$ & $\sigma_1$ \\
\hline
\end{tabular}
\end{center}
\end{table*}

The composites in Fig. \ref{fig_lumcomp} (and subsequent figures) are
plotted when at least 10 individual QSOs contribute to the spectrum.
It can be seen that as the number of QSOs is reduced the $SNR$
declines.  From this plot a number of trends can already be seen, with
the narrow \nev, \oii\ and \neiii\ showing an anti-correlation
of line strength with luminosity.  The broad emission lines of \civ,
\ciii\ and \mgii\ also show a similar correlation, appearing to
confirm previous detections of the Baldwin effect.  A further
graphical representation of these (and other) correlations is shown in
Fig. \ref{fig_ratio} which shows the luminosity segregated composites
divided by the mean composite.  This confirms that anti-correlations
with luminosity are seen for a wide variety of emission lines.  An
anti-correlation is also seen between the strength of the \caii\ H and
K absorption lines and QSO luminosity, consistent with a picture where
the host galaxy luminosity of QSOs is only weakly correlated with QSO
luminosity.  Finally we note that the Balmer series (in particular
\hb\ and \hg) appears to show a positive correlation with luminosity,
in contrast to the other emission lines.  We will analyse these
apparent correlations in a quantatative manner below.

In Fig. \ref{fig_lzcomp} we show examples of the composites divided
into absolute magnitude and redshift bins.  These allow us to decouple
the effects of redshift and luminosity.  Figs. \ref{fig_lzcomp}a and b
show composite spectra with a fixed luminosity over a range of
redshifts.  There is no obvious evidence for emission features varying
with redshift in these plots.  Figs. \ref{fig_lzcomp}c and d show
composites in a fixed redshift interval over a range in luminosity.
In this case we do see an apparent correlation between luminosity and
some lines (\civ, \nev, \oii, \neiii, \caii~K), with the lines is
question becoming weaker with increasing luminosity.  To investigate
the nature of these correlations, in particular whether they are
primarily a function of luminosity or redshift, we will carry out
detailed fitting of the spectral features, followed by a correlation
analysis.

\subsection{Line Fitting Procedure}

\begin{figure*}
\centering
\centerline{\psfig{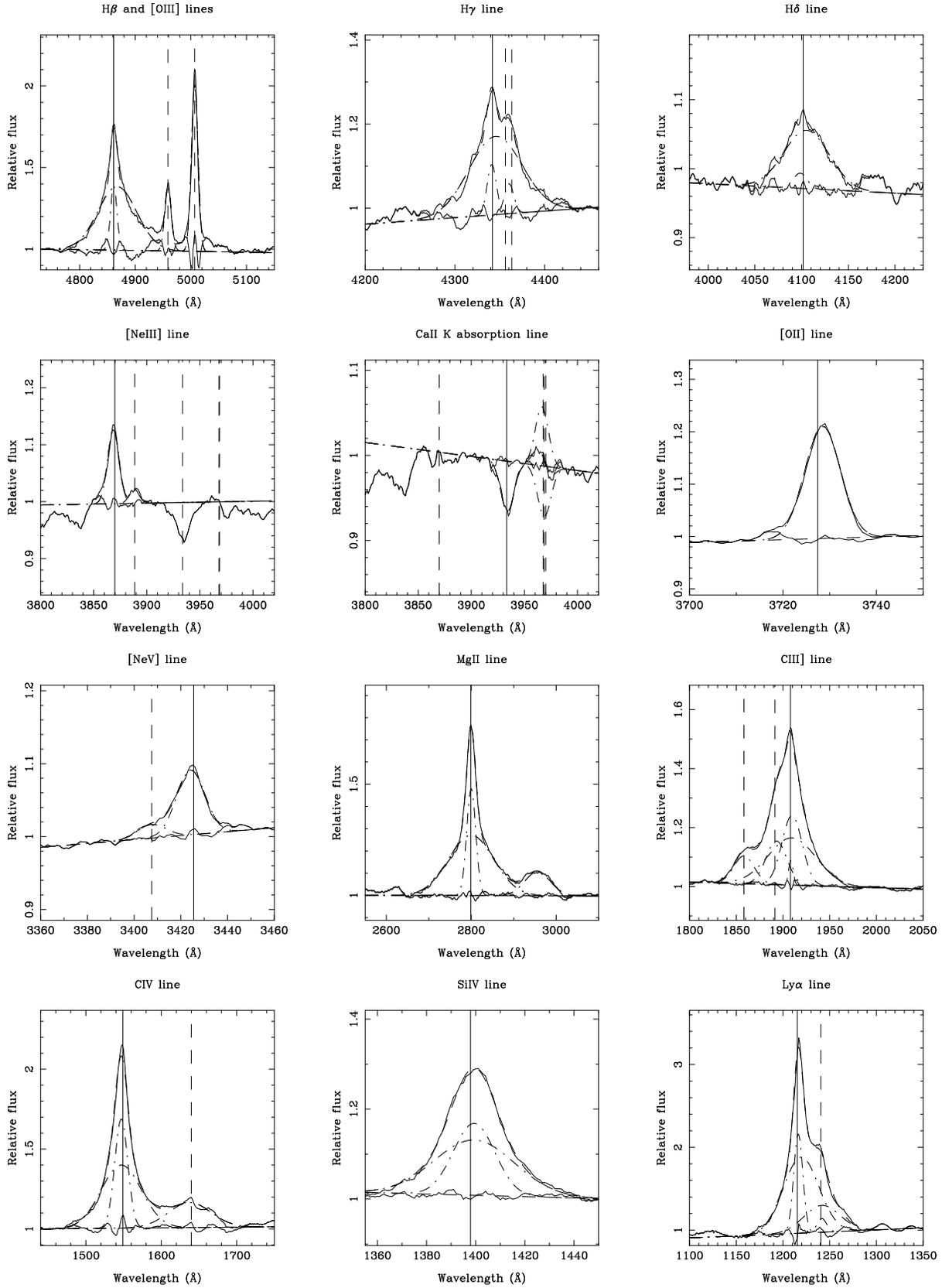}}
\caption{Example line fits.  We plot each line or complex of lines fit
in our anaylsis for the case of the total composite spectrum.  Shown
is the data (solid line), the individual Gaussian components
(dot-dashed lines), the sum of the Gaussian components (dashed line)
and the residuals after subtracting the fit (solid line).  In most
cases the dashed line denoting the total fit is hardly visible over
the data.  The vertical lines indicate the wavelengths of the primary
component (solid line) and secondary contaminating components (dashed
lines).} 
\label{fig_fits}
\end{figure*}

 The composite spectra exhibit a number of spectral features, both in
emission and absorption, which their high $SNR$ allow us to
fit. Twelve of these features, including 3 narrow (forbidden) lines, 7
broad (permitted) emission lines, one semi-forbidden line (\ciii)
and one absorption feature (\cak), were selected for
detailed study. These features were chosen because they exhibit large
equivalent widths (e.g. \la, \civ) and, in the case of the narrow
emission lines, are relatively free from contamination by other
emission lines.

The local pseudo-continuum on either side of each spectral feature was
fitted with a straight line, using a linear least-squares method, and
subtracted from the spectrum. This continuum was by no means the
'true' continuum as the emission lines in QSOs often lie on top of
other emission lines, in particular broad \feii\ features. It was,
however, relatively flat and close to the feature of interest.  The
majority of the strong emission lines in QSO spectra are blended with
other, weaker, emission lines, usually from different
elements. Additionally, permitted emission lines such as the Balmer
series often exhibit both a broad and a narrow component which are
emitted from physically distinct regions.  It is therefore necessary
to model and remove the contribution from these different lines to
obtain an accurate measurement of the line widths and equivalent
widths.

The overlapping lines contributing to each spectral feature were
modelled using multi-component Gaussian fits as listed in Table
\ref{table_fits}. We note, however, that
assuming a Gaussian form for the features in our spectra may be a
gross over-simplification, and future work will endevour to define
non-parametric measurements of line properties as well as these
Gaussian fits. Each component was fitted with a Gaussian of the form 
\begin{equation}
F(\lambda)=ae^{-\frac{(\lambda_{\rm c}-\lambda)^2}{\sigma^2}},
\end{equation}
where $a$ is the peak emission, $\lambda_{\rm c}$ is the wavelength of
the peak emission and $\sigma$ is the width of the line.  When
possible the number of independent parameters in the model was reduced
by linking some of them together. For example, since the
\oiii~\lam\lam5007,4959 and narrow \hb\ emission arises from the same
region of the QSO (the narrow line region) it is reasonable to assume
that the emitting gas will have similar velocity shifts and
dispersions.  The central wavelengths and line widths of the \oiii\
\lam4959 and narrow \hb\ emission were therefore tied to those of the
\oiii\ \lam5007. Columns 8, 9 and 10 in Table \ref{table_fits} show
how features were tied together.  For example, all the components of
the broad \hb\ line are free, while the line centres ($\lambda_{\rm
c}$) and widths ($\sigma$) for the two \oiii\ lines and the narrow \hb\
line are tied together.

The narrow emission lines were modelled as single Gaussians and were
restricted to velocity widths $<1500$~\kms. Adequate fits to the
\caii\ absorption feature and the broad Balmer emission lines (\hb,
\hg\ and \hd) were also obtained using a single Gaussian, although
there is some evidence (see Fig. \ref{fig_fits}) from these high SNR
spectra that the broad \hb\ has an asymmetric non-Gaussian
profile. The broad UV lines, i.e. from \mgii\ \lam2798
blue-ward, display emission line profiles with very broad bases which
cannot be adequately modelled by a single Gaussian. They were
therefore fitted with two components; a very broad Gaussian (FWHM
$\sim10000$~\kms) and a narrower component (FWHM
$\sim2000-4000$~\kms). This narrower component is not believed to be
emission from the narrow line region since its velocity dispersion is
much larger than that measured in the narrow lines (e.g. \oii\
\lam3727 and \oiii\ \lam5007) which is typically $\sim$ 800\kms.  When
two broad components were fitted to a broad emission line, the central
wavelengths of the Gaussians were tied together as we found no
evidence for a velocity shift between the components.  Systematic
shifts between different components, have been seen by other authors
\cite{bwfs94}, and potential line shifts will be investigated in
detail by Corbett et al. (in preparation).  The only line in which the
broad components were not tied was \la, because absorption to the blue
side of the line resulted in an asymmetric profile and it was
necessary to allow a velocity shift between the two components to fit
the line profile.  Previous studies \cite{wills93,bwss94,bwfs94} have
highlighted the fact that the broad line region can be well described
by two components, often described as the intermediate line region and
the very broad line region.  It is clear that the broad UV lines in
our composite spectra show these two components, however we reserve
detailed discussion of line shapes for the forthcoming paper, Corbett
et al.  In all cases the best fit to the spectral
feature was found using $\chi^{2}$ minimization techniques. 

The broad Balmer emission line \hg\ proved difficult to de-blend as it
is contaminated by emission from both [\feii] \lam4358 and \oiii\
\lam4363 as well as narrow \hg\ emission. Since the [\feii] and \oiii\ 
emission are within 6\AA\ of each other they are not resolved in the
2dF spectra and were therefore modelled as single narrow component
centered between the two lines. The fit was further constrained by
fixing the velocity width of the narrow \hg\ and the combined [\feii]
and \oiii\ lines to that obtained for the \oiii\ \lam5007 emission. 
It was not possible to de-blend the \oiv] $\sim\lambda$ 1402 multiplet
emission from the \siiv\ \lam\lam1393,1402 emission and hence the
equivalent width calculated for \siiv\ also contains emission from
\oiv].
 
Once the spectral feature had been modelled, the fits to the
contaminating line emission were subtracted, leaving only the line of
interest. The total flux in the line was measured by integrating the
flux over a wavelength range defined as $\lambda_{\rm
c}\pm1.5\times{\rm FWHM}$, where $\lambda_{\rm c}$ is the central
wavelength and the FWHM is that of the broadest Gaussian component
fitted to the line.  The equivalent width of the emission (or
absorption) was defined as
\begin{equation}
\ew=\frac{F_{\rm line}}{F_{\rm cont}} {\rm \AA},
\end{equation}
where $F_{\rm line}$ is the integrated flux in the emission/absorption
line and $F_{\rm cont}$ is the continuum flux measured in a 1\AA\ bin
about the central wavelength of the fit to the line. By using the
integrated residual flux in the spectral feature rather than the
Gaussian fit to calculate $F_{\rm line}$, we avoid introducing errors
due to the fact that the line emission may not be perfectly fit by a
Gaussian (e.g. \hb). There is, however, an uncertainty in $F_{\rm
line}$ due to the modelling and subtraction of the contaminating line
emission and continuum which we have taken into account when
calculating the errors in $\ew$.  In general, the multi-component fits
to the \ciii\ line were degenerate, and so we also calculate the
equivalent width of the total spectral feature, which is used in the
analysis below.

\begin{table*} 
\caption{Spearman rank correlation coefficients for correlations of 
$\log(\ew)$ with $\mb$ and $\log(1+z)$.  For each line tested we give
the number of points correlated, $N$, the Spearman rank coefficient,
$\rho$ and the probability of the null hypothesis, $P$.  Full,
bivariate coefficients and probabilities are given first for the
correlations with $\mb$ and $\log(1+z)$, then we list the partial
correlation values.}
\label{table_corr}
\begin{center}
\begin{tabular}{lrrrrrrrrr}
\hline
&&\multicolumn{4}{c}{bivariate correlations}&\multicolumn{4}{c}{partial correlations}\\
&&\multicolumn{2}{c}{$\log\ew$ vs. $\mb$}
&\multicolumn{2}{c}{$\log\ew$ vs. $\log(1+z)$} &\multicolumn{2}{c}{$\log\ew$
vs. $\mb$}&\multicolumn{2}{c}{$\log\ew$ vs. $\log(1+z)$}\\ 
Line     & $N$ &\centrho &  \centp    &\centrho &  \centp    &\centrho &   \centp   &\centrho &  \centp   \\    
\hline
\la          &  19 &  -0.476 &  3.956E-02 &   0.709 &  6.757E-04 &  -0.209 &  4.121E-01 &   0.621 &  4.908E-03\\
\nv          &  19 &  -0.119 &  6.265E-01 &  -0.059 &  8.110E-01 &  -0.170 &  5.055E-01 &  -0.135 &  5.975E-01\\
\siiv        &  30 &   0.184 &  3.303E-01 &  -0.134 &  4.811E-01 &   0.140 &  4.715E-01 &  -0.059 &  7.642E-01\\
\civ         &  34 &   0.816 &  4.058E-09 &  -0.281 &  1.080E-01 &   0.813 &  4.746E-10 &   0.258 &  1.482E-01\\
\ciii+\aliii &  49 &   0.574 &  1.650E-05 &  -0.646 &  5.436E-07 &   0.313 &  2.987E-02 &  -0.465 &  7.267E-04\\
\mgii        &  35 &   0.493 &  2.598E-03 &  -0.178 &  3.058E-01 &   0.626 &  4.236E-05 &   0.471 &  4.369E-03\\
\nev         &  23 &   0.899 &  5.625E-09 &  -0.720 &  1.060E-04 &   0.803 &  1.418E-06 &   0.330 &  1.355E-01\\
\oii         &  30 &   0.913 &  2.153E-12 &  -0.634 &  1.670E-04 &   0.917 &  1.119E-15 &   0.658 &  5.756E-05\\
\neiii       &  24 &   0.334 &  1.112E-01 &  -0.049 &  8.196E-01 &   0.598 &  2.052E-03 &   0.528 &  8.692E-03\\
\hd          &  22 &  -0.240 &  2.821E-01 &  -0.008 &  9.702E-01 &  -0.504 &  1.862E-02 &  -0.457 &  3.648E-02\\
\hg          &  22 &  -0.634 &  1.529E-03 &   0.298 &  1.786E-01 &  -0.756 &  2.876E-05 &  -0.588 &  4.189E-03\\
\hb          &  17 &  -0.711 &  1.382E-03 &   0.118 &  6.507E-01 &  -0.897 &  1.504E-07 &  -0.781 &  1.553E-04\\
\oiii        &  17 &  -0.179 &  4.920E-01 &  -0.012 &  9.628E-01 &  -0.265 &  3.280E-01 &  -0.199 &  4.675E-01\\
\cak         &  17 &   0.866 &  7.118E-06 &  -0.603 &  1.041E-02 &   0.939 &  4.146E-10 &   0.838 &  1.226E-05\\
\hline
\end{tabular}
\end{center}
\end{table*}

\subsection{Correlation analysis}

Once the widths, equivalent widths and line centres were measured for
all lines fitted above, we tested the data for correlations between
the line parameters and redshift and luminosity.  In this paper we
report the results for $\ew$.  Discussion of line widths and centres
will be reported elsewhere.  We have carried out  non-parametric rank
correlation analysis, deriving the Spearman rank-order correlation
coefficient, $\rho$.  We {\it a priori} select 99\% to be confidence
level at which we will claim significant correlations.  Specifically
we will test for an $\ew-z$ correlation by correlating $\log\ew$ with
$\log(1+z)$, as evolutionary parameters for QSOs are generally an
approximate power law in $(1+z)$, in particular, QSO luminosity
evolution \cite{2qzpaper1}.  In testing for $\ew-\mb$ correlations we
will correlate $\log\ew$ with $\mb$.  

A particularly important issue is to deduce whether $z$ or $\mb$ is
the primary parameter with which $\ew$ correlates.  We approach this
problem in two ways; the first is to carry out correlations in
separate $z$ or $\mb$ intervals, removing any possible spurious
correlations with the second independent variable.  The second
approach is to use partial Spearman rank correlation (e.g. Macklin
1982) to derive the correlation coefficient holding one independent
variable constant:
\begin{equation}
\rho_{\rm AX,Y}=\frac{\rho_{\rm AX}-\rho_{\rm XY}\rho_{\rm
AY}}{\sqrt{(1-\rho_{\rm XY}^2)(1-\rho_{\rm AY}^2)}},
\end{equation}
where $X$ and $Y$ are two independent variables (e.g. $z$ and $\mb$)
and $A$ is the dependent variable (e.g. $\ew$).  $\rho_{\rm AX}$,
$\rho_{\rm AY}$ and $\rho_{\rm XY}$ are the Spearman correlation
coefficients for the separate correlations between two variables.  The
significance of $\rho_{\rm AX,Y}$ is given by
\begin{equation}
D_{\rm AX,Y}=\frac{\sqrt{N-4}}{2}\ln\left(\frac{1+\rho_{\rm
AX,Y}}{1-\rho_{\rm AX,Y}}\right),
\end{equation}
which is distributed normally about zero with unit variance
\cite{m82}, where $N$ is the size of the sample.  In using this
partial rank correlation approach we are testing the null hypotheses
that i) the $\ew-z$ correlation arises entirely from the $\ew-\mb$
and $\mb-z$ correlations, and ii) the $\ew-\mb$ correlation arises
entirely from the $\ew-z$ and $\mb-z$ correlations.  If the
coefficients for the $\ew-z$ correlation are larger than those for
the $\ew-\mb$ correlation, this would imply that $\ew$ is {\it
primarily} correlated with $z$.

To determine the slope of any measured correlations we also carry out
fits to the data using the non-linear Levenberg-Marquardt method.
This method was used in order to fit a power law while still properly
taking into account the errors on the $\ew$  measurements (which is
not possible in a standard linear least-squares approach).  As will be
seen below, the errors in the individual $\ew$ measurements were often
much smaller than the scatter about the best fit line.  This could
reflect the fact that parameters other than those fitted are
introducing extra dispersion and/or the power law is not a adequate
fit to the data.  To obtain a realistic error on the fitted
parameters, we repeat the fitting procedure, rescaling the errors such
that the reduced $\chi^2$ is exactly one in each case, noting the
specific cases where the data diverge significantly from a power law.

\begin{figure*}
\centering
\centerline{\psfig{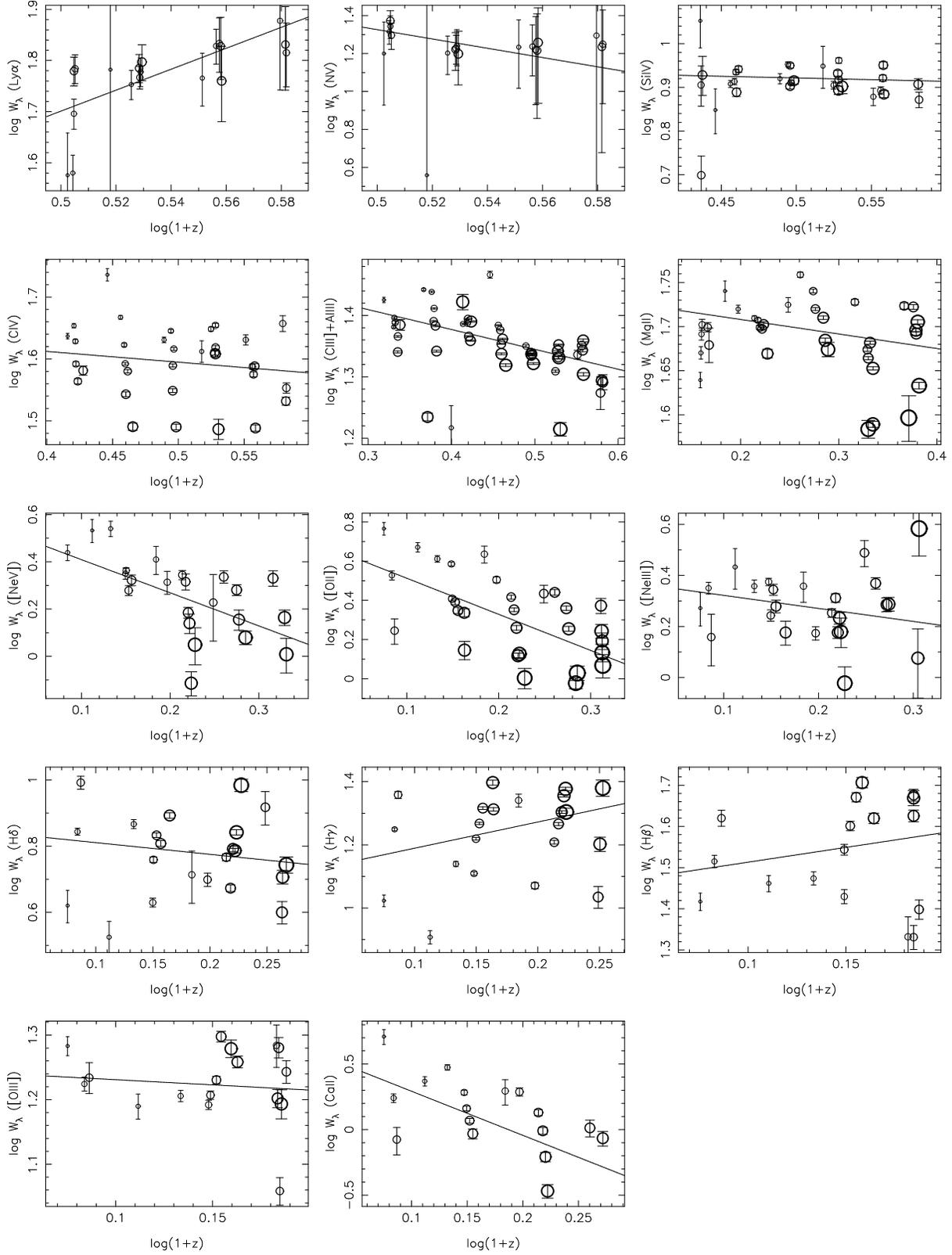}}
\caption{The correlations of $\log\ew$ with $\log(1+z)$ for each line.
The circles denote the measured values of $\log\ew$, with larger
circles indicating more luminous intervals in $\mb$.  No significant
correlations are found in individual luminosity intervals.  For every
feature, regardless of whether a significant correlation is seen or not,
the best power law fit to all the data points is shown (solid
line).}
\label{fig_zcorr}
\end{figure*}

\section{results}

We derive both the bivariate and partial Spearman rank correlation
coefficients for the correlation of $\ew$ with $z$ and $\mb$.  Our
primary aims in doing this are to i) test for the existence of any
significant correlations, and ii) determine if those correlations are
primarily with $z$ or $\mb$.  This second point is important given the
recent claim \cite{gfk01} that the Baldwin effect is primarily a
correlation with redshift.

\subsection{Bivariate and partial Spearman rank correlation}

We select the strongest and cleanest spectral features to test for
correlations.  In cases where there is significant contamination by 
other features the Gaussian fits for these have been subtracted off
the summed flux to provide a clean estimate of line flux.  This has not
been done in a few cases, where the separate components cannot be 
clearly distinguished.  In particular, we use the summed flux of all
components in the \siiv+\oiv] complex, and do not subtract off narrow 
components from \hd\ or \hg\ (including \oiii\ \lam4363).  Note, however
that we combine together broad and intermediate components of the same
line (in particular for all the broad UV lines).

\begin{figure*}
\begin{center}
\centerline{\psfig{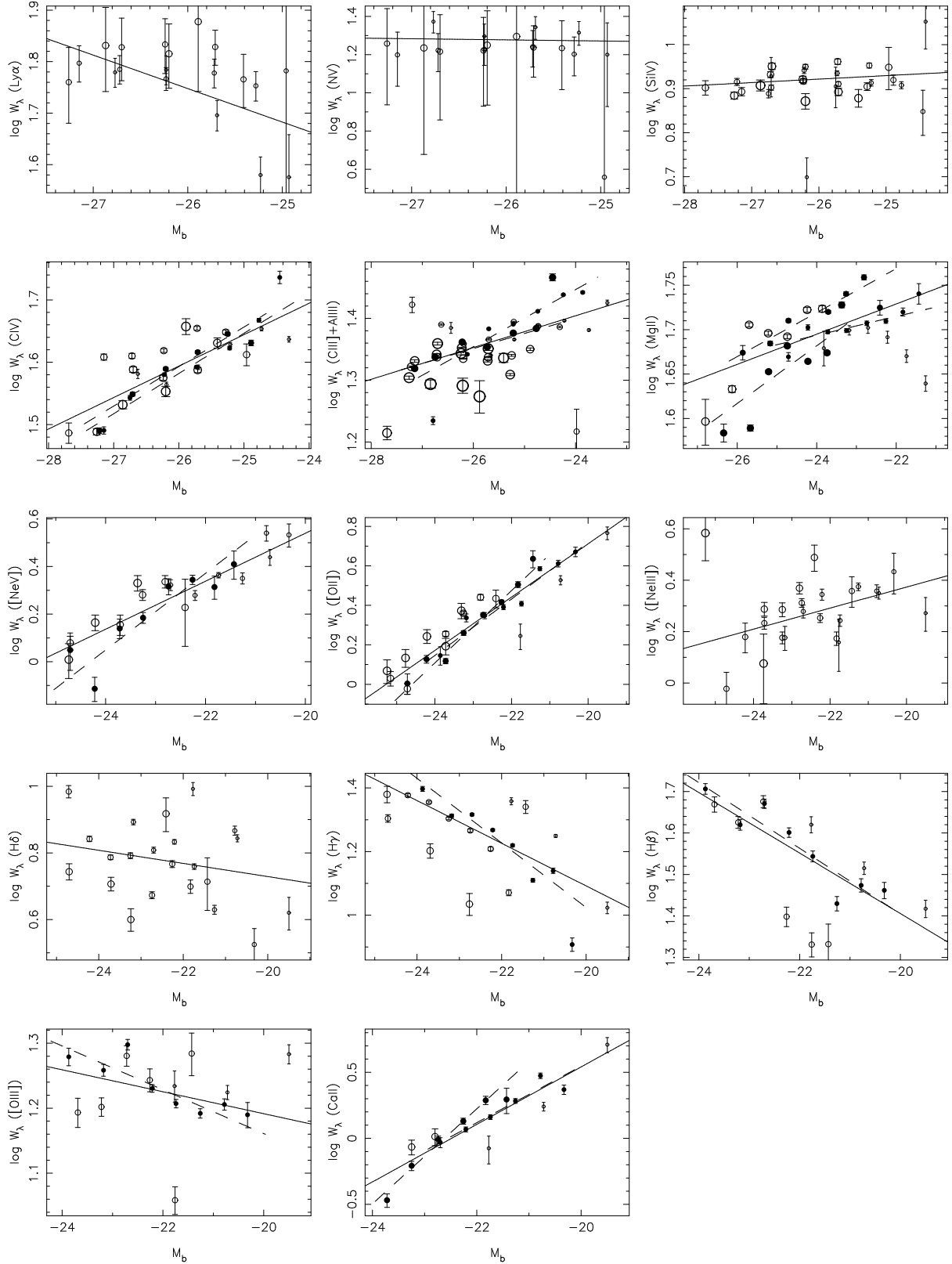}}
\caption{The correlations of $\log\ew$ with $\mb$ for each
line.  The circles denote the measured values of $\log\ew$, with
larger circles indicating higher redshift intervals.  In cases
where an individual luminosity interval shows a significant
correlation with redshift the circles are filled and the best fit
power law correlation is plotted (dashed lines).  For every feature,
regardless of whether a significant correlation is seen or not, the
best power law fit to all the data points is also shown (solid line).}
\label{fig_lumcorr}
\end{center}
\end{figure*}

We carry out the correlations first for $\log(1+z)$ and $\mb$.  The
resulting correlation coefficients are listed in Table
\ref{table_corr}.  The number of points used in the correlations
ranges from 17 to 49 with a median of 22.  Table \ref{table_corr}
first lists the full bivariate correlation coefficients and
probabilities for the $\log\ew$ correlations with $\mb$ and
$\log(1+z)$.  These do not take into account any potential spurious
correlation caused by the correlation of $\mb$ and $\log(1+z)$.  We
find that a number of lines show significant ($P<0.01$) correlations.
The \civ, \ciii+\aliii, \mgii, \nev, \oii, \cak, \hg\ and \hb\ lines
all show  significant correlations with $\mb$.  We find that less
lines, only \la, \ciii+\aliii, \nev\ and \oii, show correlations with
$\log(1+z)$.  The data (filled and open circles) and best fit
correlations (solid lines) are shown in Figs. \ref{fig_zcorr} and
\ref{fig_lumcorr}.

We then derive the partial Spearman rank correlation coefficients of
$\log\ew$ with $\mb$ and $\log(1+z)$, which are listed in the last
four columns of Table \ref{table_corr}.  In all but two cases (\la,
\ciii+\aliii) the significance of the correlation is larger for $\mb$
than $\log(1+z)$.  The strongest partial correlations with $\log(1+z)$
are for \oii\ and \caii~K which are also the lines showing the
steepest correlations with $\mb$.  This is consistent with these
$\log(1+z)$ correlations being due to the luminosity distribution of
the QSOs, {\it within} a given luminosity  range, changing with
redshift.  Our partial correlation analysis demonstrates that the
correlations seen are primarily with $\mb$ rather than redshift, in
disagreement with the previous results of Green et al. (2001).  

\subsection{The correlation of $\ew$ and redshift}

To further investigate the finding that $\ew$ primarily correlates
with $\mb$ and not $z$ we now use the separate luminosity intervals of
width $\Delta\mb=0.5$ to search for correlations with redshift.  In a
given luminosity interval there are up to eight $\Delta z=0.25$
redshift intervals sampled.  For each measured emission line we test
for a correlation between $\ew$ and $\log(1+z)$ independently within
each luminosity interval using Spearman rank correlation.  In some
luminosity intervals there may be only a small number of redshift
intervals in which a particular line is present.  This is particularly
true for lines such as \hb\ and \oiii\ near the edge of the spectrum,
which are only present in 3 redshift intervals.
In the Spearman rank correlation analysis we limit ourselves to
examining  luminosity intervals which contain at least 5 separate
measurements.  This is because the significance of $\rho$ is derived
from $t=\rho\sqrt{(N-2)/(1-\rho^2)}$ which is approximately
distributed as Student's distribution.  However this breaks down for
small $N$, as it predicts zero probability for $\rho=\pm1$, whereas
the true likelihood of this occurring is ($2/N!$).  Only the \siiv,
\civ, \ciii+\aliii\ and \mgii\ lines have 5 or more measured equivalent
widths in a given luminosity interval, hence only these lines are
sensitive to tests for $\log\ew$ vs. $\log(1+z)$ correlations in each
luminosity interval.  No significant correlations are found for any of
these lines, supporting the above finding that the correlations are
primarily driven by luminosity.  Fig. \ref{fig_zcorr} shows the
distribution $\log\ew$ vs. $\log(1+z)$ for all the lines.  The symbols
(circles) are larger for brighter luminosity intervals, and in a
number of cases (e.g. \oii) we see that the correlation could
potentially be due to intrinsically fainter (small circles) sources
having larger equivalent widths.  The above partial correlation
analysis confirms this impression.

\begin{table*} 
\caption{Best fit linear correlations with Spearman rank correlation
coefficients and probabilities.  First we list the best fits for all
redshift intervals combined.  Then we list correlations from those
lines which show significant correlations in individual redshift
intervals.  $N$ is the number of points used in the correlation
analysis.  $A$ and $B$ are the intercept and gradient for the best fit
line, such that $\log\ew=A+B\mb$.  $\delta A$ and $\delta B$ are the
errors on the intercept and slope.  $\rho$ and $P$ are the Spearman
rank coefficient and probability respectively.  Probabilities marked
by an asterisk ($^*$) have $\rho=1$ (a perfect correlation).  These
probabilities have been corrected to $P=(2/N!)$.}
\label{table_corrfit}
\begin{center}
\begin{tabular}{lrrrrrrrr}
\hline
Line     & N
& \multicolumn{1}{c}{redshift} & \multicolumn{1}{c}{$A$} &
\multicolumn{1}{c}{$\delta A$} & \multicolumn{1}{c}{$B$} &
\multicolumn{1}{c}{$\delta B$} &
\multicolumn{1}{c}{$\rho$} & \multicolumn{1}{c}{$P$}\\
\hline
\la          &  19 &  0.00 -- 4.00 &   0.059 &   0.531 &  -0.065 &   0.020 &  -0.476 &  3.956E-02 \\
\nv          &  19 &  0.00 -- 4.00 &   1.124 &   1.004 &  -0.006 &   0.039 &  -0.119 &  6.265E-01 \\
\siiv        &  30 &  0.00 -- 4.00 &   1.127 &   0.173 &   0.008 &   0.007 &   0.184 &  3.303E-01 \\
\civ         &  34 &  0.00 -- 4.00 &   2.905 &   0.154 &   0.050 &   0.006 &   0.816 &  4.058E-09 \\
\ciii+\aliii &  49 &  0.00 -- 4.00 &   2.021 &   0.097 &   0.026 &   0.004 &   0.574 &  1.650E-05 \\
\mgii        &  35 &  0.00 -- 4.00 &   2.103 &   0.109 &   0.017 &   0.005 &   0.493 &  2.598E-03 \\
\nev         &  23 &  0.00 -- 4.00 &   2.557 &   0.263 &   0.101 &   0.012 &   0.899 &  5.625E-09 \\
\oii         &  30 &  0.00 -- 4.00 &   3.409 &   0.222 &   0.135 &   0.010 &   0.913 &  2.153E-12 \\
\neiii       &  24 &  0.00 -- 4.00 &   1.192 &   0.325 &   0.041 &   0.015 &   0.334 &  1.112E-01 \\
\hd          &  22 &  0.00 -- 4.00 &   0.334 &   0.370 &  -0.020 &   0.016 &  -0.240 &  2.821E-01 \\
\hg          &  22 &  0.00 -- 4.00 &  -0.249 &   0.257 &  -0.067 &   0.011 &  -0.634 &  1.529E-03 \\
\hb          &  17 &  0.00 -- 4.00 &  -0.053 &   0.353 &  -0.073 &   0.016 &  -0.711 &  1.382E-03 \\
\oiii        &  17 &  0.00 -- 4.00 &   0.854 &   0.235 &  -0.017 &   0.011 &  -0.179 &  4.920E-01 \\
\cak         &  17 &  0.00 -- 4.00 &   4.895 &   0.488 &   0.218 &   0.022 &   0.866 &  7.118E-06 \\
\hline
\civ         &   7 &  1.75 -- 2.00 &   3.260 &   0.196 &   0.065 &   0.008 &   1.000 &  3.968E-04$^*$ \\
\civ         &   6 &  2.00 -- 2.25 &   3.240 &   0.248 &   0.063 &   0.010 &   0.943 &  4.805E-03 \\
\ciii+\aliii &   7 &  1.25 -- 1.50 &   2.556 &   0.163 &   0.046 &   0.007 &   1.000 &  3.968E-04$^*$ \\
\ciii+\aliii &   7 &  1.75 -- 2.00 &   1.984 &   0.174 &   0.024 &   0.007 &   0.964 &  4.541E-04 \\
\mgii        &   8 &  0.50 -- 0.75 &   1.948 &   0.072 &   0.011 &   0.003 &   0.929 &  8.630E-04 \\
\mgii        &   7 &  0.75 -- 1.00 &   2.314 &   0.104 &   0.025 &   0.004 &   0.893 &  6.807E-03 \\
\mgii        &   7 &  1.00 -- 1.25 &   2.461 &   0.314 &   0.032 &   0.013 &   0.893 &  6.807E-03 \\
\nev         &   8 &  0.50 -- 0.75 &   3.915 &   0.621 &   0.161 &   0.027 &   0.905 &  2.008E-03 \\
\oii         &   8 &  0.25 -- 0.50 &   3.574 &   0.510 &   0.143 &   0.023 &   1.000 &  4.960E-05$^*$ \\
\oii         &   8 &  0.50 -- 0.75 &   4.500 &   0.329 &   0.183 &   0.014 &   0.976 &  3.314E-05 \\
\hg          &   8 &  0.25 -- 0.50 &  -1.006 &   0.343 &  -0.102 &   0.015 &  -0.952 &  2.604E-04 \\
\hb          &   8 &  0.25 -- 0.50 &  -0.178 &   0.319 &  -0.079 &   0.014 &  -0.905 &  2.008E-03 \\
\oiii        &   8 &  0.25 -- 0.50 &   0.487 &   0.205 &  -0.034 &   0.009 &  -0.905 &  2.008E-03 \\
\cak         &   6 &  0.25 -- 0.50 &   4.823 &   0.794 &   0.214 &   0.037 &   0.943 &  4.805E-03 \\
\cak         &   6 &  0.50 -- 0.75 &   8.139 &   0.651 &   0.360 &   0.029 &   1.000 &  2.778E-03$^*$ \\
\hline
\end{tabular}
\end{center}
\end{table*}

\subsection{The correlation of $\ew$ and $\mb$}

We now correlate $\ew$ with $\mb$ in separate redshift intervals of
$\Delta z=0.25$.  Again, only intervals with 5 or more $\ew$
measurements are tested for correlations, however each line has at
least one redshift interval with 5 or more points.
Fig. \ref{fig_lumcorr} shows the results of this analysis.  We find
that all of the lines which show significant correlations over the
entire redshift interval also show significant correlations in at
least one individual redshift interval.  In fact, \civ, \ciii+\aliii,
\mgii, \oiii\ and \caii~K all show significant correlations with
$\log\ew$ in two or more redshift intervals. 

We note the odd behaviour of the \mgii\ equivalent width in the lowest
redshift interval, exhibiting a trend with luminosity in the opposite
sense to the other redshift ranges.  It is difficult to ascribe this
to a selection effect; any Malmquist bias in the measurement of the
equivalent width (occurring when a particular line is the dominant or
only line responsible for the identification of a quasar at a
particular redshift e.g. \mgii) would likely give rise to the opposite
effect i.e. a tendency to over-estimate the mean equivalent width at
the faint magnitudes (lowest luminosities) in the sample.

Indeed, some Malmquist bias may be present in the data, although we
have confirmed that the results presented here are robust against the
inclusion/exclusion of the faintest 0.5\,mag interval in absolute
magnitude.  Moreover, the fact that a range of slopes are found for
the correlation between equivalent width and $\mb$ (both positive and 
negative) further suggests that any Malmquist bias plays a small role.

In Table \ref{table_corrfit} we list the parameters of all the
significant correlations, including their significance and best fit
parameters for the fit to $\log\ew=A+B\mb$.  From Table
\ref{table_corrfit} we can see that there are significant differences
between gradients of the different lines.  The strongest correlation
is found in the \caii~K line with a gradient of $0.218\pm0.022$.  The
Balmer lines are the only ones to show a negative correlation with
$\mb$ (a positive correlation with luminosity), which confirms the
visual impression gained from Fig. \ref{fig_ratio}.  We discuss the
physical significance of these correlations below.  

\begin{table*} 
\caption{Best fit linear correlations with Spearman rank correlation
coefficients and probabilities for composites sub-divided by
luminosity only.  $N$ is the number of points used in the correlation
analysis.  $A$ and $B$ are the intercept and gradient for the best fit
line, such that  $\log\ew=A+B\mb$.  $\delta A$ and $\delta B$ are the
errors on the intercept and slope.  $\rho$ and $P$ are the Spearman
rank coefficient and probability respectively.}
\label{table_corrfitmagbin}
\begin{center}
\begin{tabular}{lrrrrrrr}
\hline
Line & N & \multicolumn{1}{c}{$A$} & \multicolumn{1}{c}{$\delta A$} &
\multicolumn{1}{c}{$B$} & \multicolumn{1}{c}{$\delta B$} &
\multicolumn{1}{c}{$\rho$} & \multicolumn{1}{c}{$P$}\\
\hline
\la          &   5 &   0.822 &   0.665 &  -0.037 &   0.026 &  -0.400 &  5.046E-01 \\
\nv          &   5 &   1.781 &   1.118 &   0.020 &   0.043 &   0.800 &  1.041E-01 \\
\siiv        &   8 &   1.118 &   0.181 &   0.008 &   0.007 &  -0.143 &  7.358E-01 \\
\civ         &   8 &   2.917 &   0.143 &   0.051 &   0.006 &   0.976 &  3.314E-05 \\
\ciii+\aliii &  10 &   2.075 &   0.080 &   0.028 &   0.003 &   0.976 &  1.468E-06 \\
\mgii        &  12 &   2.233 &   0.116 &   0.023 &   0.005 &   0.720 &  8.240E-03 \\
\nev         &  10 &   2.423 &   0.369 &   0.095 &   0.016 &   0.964 &  7.321E-06 \\
\oii         &  12 &   3.330 &   0.205 &   0.131 &   0.009 &   0.986 &  4.117E-09 \\
\neiii       &  11 &   0.984 &   0.377 &   0.031 &   0.017 &   0.645 &  3.196E-02 \\
\hd          &  11 &   0.546 &   0.265 &  -0.009 &   0.012 &  -0.491 &  1.252E-01 \\
\hg          &  13 &  -0.203 &   0.175 &  -0.064 &   0.008 &  -0.703 &  7.319E-03 \\
\hb          &  10 &  -0.066 &   0.259 &  -0.074 &   0.012 &  -0.927 &  1.120E-04 \\
\oiii        &  10 &   0.862 &   0.237 &  -0.017 &   0.011 &  -0.527 &  1.173E-01 \\
\cak         &   9 &   4.703 &   0.492 &   0.208 &   0.023 &   0.983 &  1.936E-06 \\
\hline
\end{tabular}
\end{center}
\end{table*}

\begin{figure*}
\centering
\centerline{\psfig{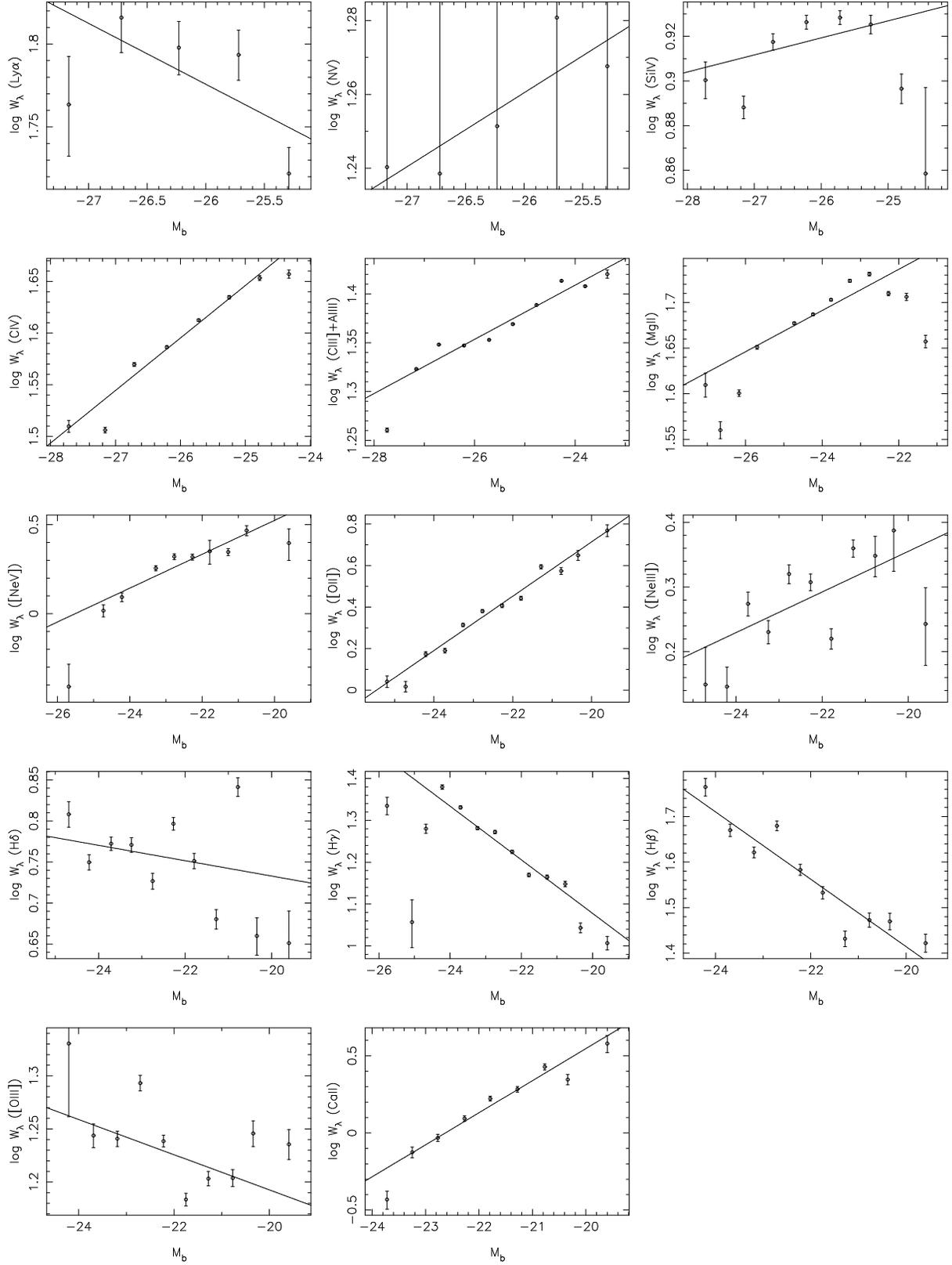}}
\caption{The correlations of $\log\ew$ with $\mb$ for each
line, measured from composites sub-divided by luminosity only.  The
best fit correlation is plotted in each case.}
\label{fig_magbincorr}
\end{figure*}

\subsection{The correlation of $\ew$ with $\mb$ in luminosity only
divided bins}

Given that the above analysis appears to suggest that the dominant
correlation is with $\mb$ we now derive correlations between $\log\ew$
with $\mb$ for the composites sub-divided only on the basis of
luminosity (Fig. \ref{fig_lumcomp}).  This can potentially reduce the
noise and scatter in the correlations if $\ew$ is truly only
correlated with $\mb$.  The resulting correlations are shown in
Fig. \ref{fig_magbincorr}.  The correlation coefficients and best fit
parameters are listed in Table \ref{table_corrfitmagbin}.  We see
evidence of significant correlations in many lines, with some
exceptions.  The lines which do not show correlations are \la, \nv,
\siiv, \neiii, \hd\ and \oiii.  We also note that in some cases, most
notably \civ, \ciii+\aliii\ and \mgii, the dispersion about the best
fit correlation is much larger than would be expected give the errors
on individual points.  This suggests that other parameters may cause
extra dispersion in the relation, or that a simple power law fit is
not actually a good description of the underlying physics. 

\section{Discussion}

We now attempt to understand the above measured correlations in the
context of simple physical models for AGN emission and host galaxy
properties.  We will start by considering the host galaxy, and then
look at the narrow line and broad line regions in turn.

\subsection{Host galaxy properties}

The \cak\ absorption line is possibly the most simple to interpret, as
it can only be due to the stars present in the host galaxy of the QSO.
We clearly see that as the AGN luminosity increases, the strength of
the \cak\ declines, consistent with a picture in which the host galaxy
does not increase in luminosity as fast as does the AGN.  If the host
galaxy was constant in luminosity, the slope of the correlation
between $\log\ew$ and $\mb$ would be 0.4.  Our best fit slope is
$0.208\pm0.023$, which is therefore consistent with host galaxy
luminosity increasing slowly.  In this analysis we make the assumption
that the average spectral properties of the AGN host galaxies do not
change significantly with luminosity (or redshift). We also assume
that there is no significant aperture effect introduced by the
2-arcsec diameter of the 2dF fibres used to obtain the spectra.  Based
on the imaging results of Schade, Boyle \& Letawsky (2000), we expect
that galaxy  hosts for QSOs in this luminosity range ($-21>M_B>-24$)
will be bulge-dominated with effective radii ranging from 1 -- 2\,kpc
for QSOs at  $z\sim0.15$ and 3 -- 6\,kpc for QSOs with $z\sim0.6$.
In both cases the projected size of the bulges are is approximately
the same  size on the sky ($\sim$ 0.75 -- 1.5-arcsec diameter) and
significantly less  than the fibre diameter.

\begin{figure}
\centering
\centerline{\psfig{file=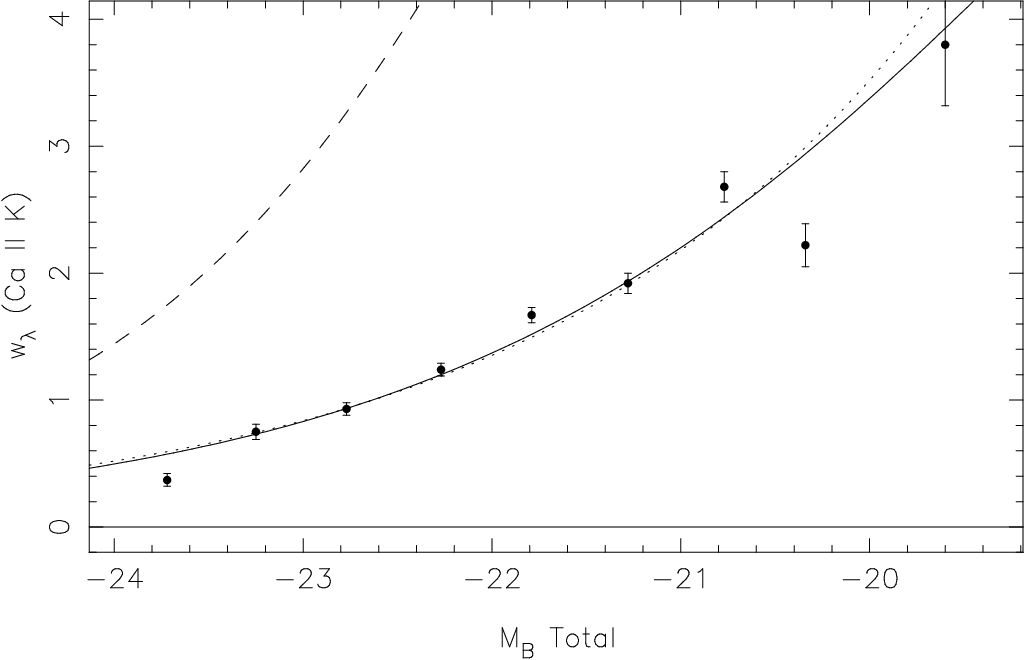,width=8cm}}
\caption{The relationship between $\ew$(\cak) and $\mb$ for
composites sub-divided by  luminosity only.  Three models are shown,
$\mgal=A+B\mqso$ (solid line), $\mgal=A+B\mtot$ (dotted line) and
$\mgal=A+B\mqso$ with parameters set from Schade et al. (2000) (dashed
line).}
\label{fig_ca2}
\end{figure}

If we also assume that the majority of the continuum emission is due
to the QSO, we can derive a simple relation for the expected
correlation between $\ew$ and $\mb$.  We can set $L_{\rm line}\propto
\lgal\propto\lqso^\alpha$, that is,  the QSO luminosity is
proportional to the host galaxy luminosity to some power.  If
$\alpha=0$ then the host galaxy has a constant luminosity.  Converting
to magnitude and $\log\ew$ we then find
\begin{equation}
\log\ew=0.4(1-\alpha)\mb+Const.
\end{equation}
Taking our fitted value for the slope of the correlation we find that
$\alpha=0.48\pm0.06$.  This then implies that QSO and host galaxy
luminosity are correlated.  This has in fact been found from direct
imaging studies of QSO host galaxies.  Schade et al. (2000) find
$\lgal\propto\lqso^{0.21}$ for a sample of low redshift X-ray selected
AGN, with large scatter.  This is a somewhat shallower slope than we
find, however, we are in fact deriving the slope for the relation
$\lgal\propto\ltot^\alpha$, not $\lgal\propto\lqso^\alpha$.  We can
instead construct a model which assumes an exact power law correlation
between $\lgal$ and $\lqso$.  However this does depend on us having
some knowledge of the expected equivalent width for \cak\ in the host
galaxy, without the AGN component.  We cannot make an accurate
assessment of the spectral properties of the host galaxy, as we have
only fit one feature.  Instead we use a \cak\ line strength derived
from the mean galaxy spectrum in the 2dF Galaxy Redshift Survey
\cite{cosspec}.  This has an equivalent width for \cak\ of
$7.3\pm0.2$\AA.  The maximum $\ew$ found for \cak\ in our analysis is
$\sim4-5$\AA, thus at the faintest luminosities the the host galaxy
could be contributing a significant fraction of the continuum.  If the
equivalent width of the line in the galaxy spectrum is $\ewg$, then
\begin{equation}
\ew=\frac{\ewg}{\ltot/\lgal}=\frac{\ewg}{10^{-0.4(\mtot-\mgal)}},
\label{eq_ca2}
\end{equation}
where $\mtot$ and $\mgal$ are the absolute B-band magnitudes from the
total (AGN+host) and host components.  Assuming a linear relation
between $\mgal$ and $\mqso$ such that $\mgal=A+B\mqso$ implies that
\begin{equation}
\mgal=A-2.5B\log(10^{-0.4\mtot}-10^{-0.4\mgal})
\end{equation}
which can be solved numerically for $\mgal$, and substituted into
Eq. \ref{eq_ca2}.  The resulting best fit is
\begin{equation}
\mgal=(-11.10\pm0.97)+(0.417\pm0.045)\mqso,
\end{equation}
which is shown in Fig. \ref{fig_ca2} (solid line).  This fit is very
similar to the previous power law fit (dotted line), only diverging at
faint magnitudes.  Also plotted is the relation found by Schade et
al. (2000) of $\mgal=-17.25+0.21\mqso$, this is clearly discrepant
with our data.  Reducing the intrinsic strength of \cak\ to
$\sim4.5$\AA\ does not remove this discrepancy.  We note that Schade et
al. have demonstrated that for low redshift AGN ($z\sim0.1$) at about
$L^*$ the host galaxies are in almost all respects no different to
normal galaxies.  The one difference found is a bias towards
spheroidal morphologies.  This could imply that the stellar
populations in AGN host galaxies are older than in average galaxies,
but this is by no means certain given that AGN activity could also be
accompanied by enhanced star formation. 

It is also possible to use the measured \cak\ line together with a
mean galaxy spectrum to determine the expected spectral properties of
other features not produced by the AGN, in particular, the narrow
forbidden oxygen emission lines.  In the mean 2dF Galaxy Redshift
Survey spectrum the flux emitted  in \oiii~\lam5007 is about a factor
of two lower that the flux absorbed in \cak.  However, even in our
faintest composites where the host galaxies contribute the most, the
\oiii\ line has an $\ew$ a factor of over 4 greater than \cak.  Thus,
assuming that the host galaxy SED is similar to that of a normal
galaxy implies that the \oiii\ emission is coming from the AGN, rather
than the host galaxy.  

The situation regarding the  \oii~\lam3727 line is different. In the
mean galaxy spectrum the line has a relative flux which is similar to
that in the \cak\ line, and has an equivalent width of
$\sim10.7\pm0.3$\AA.  Comparing the correlations of \cak\ and \oii\ we
find that close to the faint end of the distribution, at $\mb=-20$,
the \oii\ $\ew$ is 70--100\% of what would be predicted from the host
galaxy.  As the \oii\ correlation with $\ew$ is flatter than that of
\cak\, the simple assumption of a constant host galaxy SED would
predicit that the increased fraction of the \oii\ flux is emanating
from the AGN at higher luminosities ($\sim65$\% at
$\mb=-24$). However, the difference is not as large as in the case of
the \oiii\ line and  it is possible to speculate that the \oii\ line
is formed in powerful star-forming regions and  the star-formation
rate in the host galaxy is increasing with AGN luminosity. In this
case, all the \oii\ emission could be due to the host galaxy.  The
\oii\ line would then provide a very useful diagnostic tool for the
study of star formation in high redshift QSOs.  A further test of this
would be to investigate the velocity distribution of the different
narrow lines.  If, for example, the \oii\ line has a significantly
lower velocity dispersion, more consistent with typical galaxies, than
\oiii\, this would be good evidence of for the \oii\ being due mostly
to the host galaxy. Here we proceed by assuming that a major fraction
of the \oii\ line emission originates in the NLR of the AGN. The
former possibility of the starburst origin  will be investigated
further by Corbett et al. (in preparation).

\subsection{The narrow-line region}

\begin{figure}
\centering
\centerline{\psfig{file=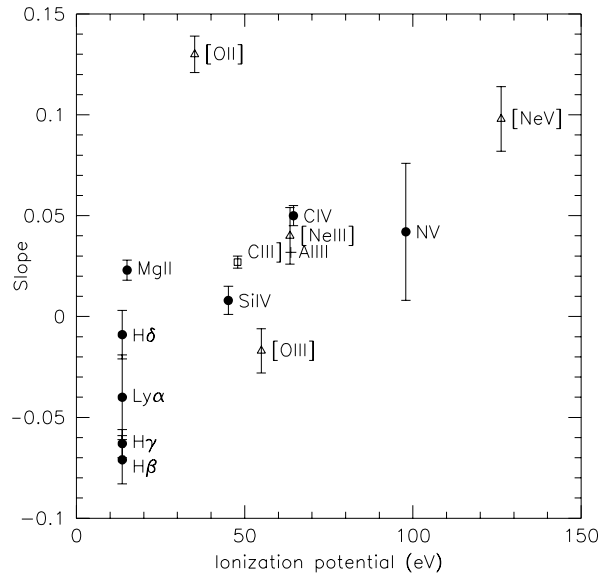,width=8cm}}
\caption{The measured slopes of the $\log\ew$ vs. $\mb$ correlation
for different lines as a function of lower ionization potential for
broad permitted lines (filled circles), narrow forbidden lines (open
triangles) and semi-forbidden \ciii\ (open square).}
\label{fig_ionslope}
\end{figure}

 The lines that are thought to be  emitted within the NLR of the AGN
are \oii, \oiii, \neiii\ and \nev\ (see however comment about \oii\ in
the previous section).  Of those, only \oii\ and  \nev\ show  clear
and significant correlations with $\mb$.  The partial correlation
anaylsis for both lines is consistent with the hypothesis that the
correlation is solely due to variations in $\mb$, with no correlation
with redshift.  The slope of the \nev\ correlation for the magnitude
only divided composites is $0.095\pm0.016$, consistent with the
correlation found in the $\mb-z$ composites, which has a slope of
$0.101\pm0.012$.  This line has the highest ionization potential (here
and below we consider the lower ionization potential; the energy
required to ionize the lower ionization ion) of any of the narrow
lines we investigate (126.21~eV). The \oii\ line has the lowest
ionization potential and the other two have intermediate values:
63.45~eV  for \neiii\  and 54.93~eV  for \oiii.  The \neiii\ line
shows only a marginal detection of a correlation (97\% significant)
and has a significantly flatter slope, $0.031\pm0.017$.  The \oiii\
line shows no evidence of a correlation and all the measured
equivalent widths in the $\mb$ composites are within a range of
$\Delta\log\ew=0.15$.  Plotting the measured slope as a function of
ionization energy for these narrow lines (open triangles in
Fig. \ref{fig_ionslope}) we see no obvious trend. However, if much of
the \oii\ emission is due to star formation, as suggested in the earlier
section, there may be a correlation in a sense that  a steeper slope
corresponds to higher ionization energy. Given the small number of
points and the additional assumption about \oii, we cannot use this
trend to infer the NLR physics.

Previous analyses have tentitively detected correlations between
narrow line strength and luminosity \cite{gfk01}.  However, these were
generally in data sets with a large fraction of non-detections, as
they used single objects instead of composite spectra.  The suggested
correlation of line intensity  with luminosity in \oiii\ demonstrated
in Fig. \ref{fig_ratio} does not appear to be borne out in the
correlation analysis involving equivalent width measurements.  The
apparent variation could be due to a real variation in the velocity
width of the \oiii\ lines while the total flux remains approximately
constant.

A simple interpretation of the \nev\ and \oii\ correlation involves
the 'disappearing NLR' model. This idea is based on the fact that
the NLR size scales with the source luminosity to some power
\begin{equation}
R_{\rm NLR} =R_0 (L/L_0)^{\beta},
\end{equation}
which is reasonable given the similar correlation found for the broad
line region (BLR) \cite{k00} and the actual observed NLR size in a
number of sources. Luminosity scaling suggests $0.5<\beta<0.7$ where
the lowest value is obtained from the similarity of the narrow line
spectrum in low and high luminosity AGN (e.g. Netzer 1990) and the
$\beta=0.7$ is the value obtained by Kaspi at al. (2000) for the
BLR. For nearby low luminosity Seyfert 1s, $R_0 \simeq 500$~pc. Thus,
the 5 magnitude range in  $\mb$ observed in our sample translates to
$R_{\rm NLR}=5-13$~kpc for the highest luminosity AGN, i.e.  the scale
of the entire galaxy, and indeed, recent HST imaging of the NLR in
several radio-loud quasars shows equation (9) to hold for R$_{NLR}$ up
to $\sim$10 kpc with $\beta$=0.5 \cite{b02}. It is therefore possible
that the NLR gas, if was there in the first place, has long left the
galaxy and most high luminosity AGN contain weak or non-existent
NLRs. Moreover, if $\beta=0.7$ as suggested here, and if the NLR
density is independent of size, we can perhaps explain the decreasing
equivalent width of \nev\ as due to a decreasing ionization parameter
with luminosity. 

The \neiii\ $\ew$ correlation with luminosity is marginal although
with the right trend (Fig. \ref{fig_magbincorr}). However, the  above
simplified model does not explain the different behaviour of the
\oiii\ line that shows no obvious correlation. We note, however, that
the \neiii, \nev\ and \oii\ lines are measured over a larger magnitude
range, compared with \oiii, because the \oiii\ lines are lost off the
red end of the spectrum at much lower redshifts (and hence lower
luminosities in a flux limited sample) than the other emission lines.
Thus, the reality of the model will have to be tested when more \oiii\
line measurements of higher luminosity AGN are available.  Needless to
say, if the model suggested here is confirmed by future observation,
it will have serious implications for searches for luminous {\it type
2} QSOs which simply may not have luminous narrow emission line
regions.

\begin{figure}
\centering
\centerline{\psfig{file=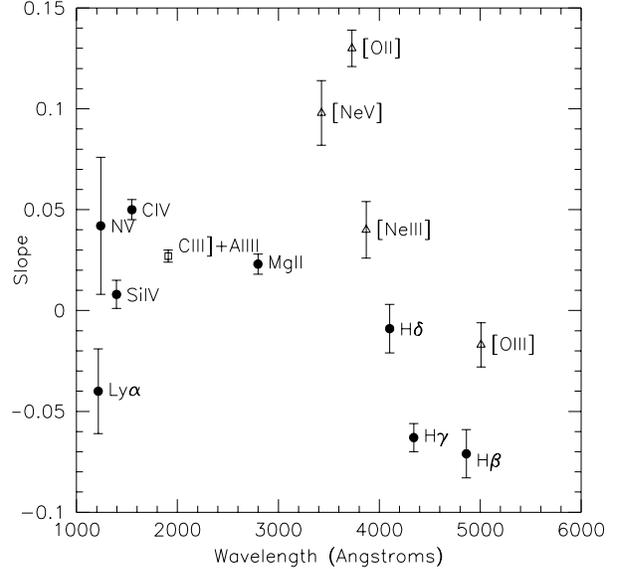,width=8cm}}
\caption{The measured slopes of the $\log\ew$ vs. $\mb$ correlation
for different lines as a function of rest wavelength for broad
permitted lines (filled circles), narrow forbidden lines (open
triangles) and semi-forbidden \ciii\ (open square).}
\label{fig_lamslope}
\end{figure}

\subsection{The broad line region}

Lastly we consider emission from the BLR.  Most of the strong emission
lines we analyze emanate from this region and  we will discuss each of
the lines in turn.

The \la\ line is the only one to show a clear correlation with redshift
(top left plot in Fig. \ref{fig_zcorr}). We see an increase in \la\
strength with redshift.  However, this is likely to be due to a
combination of increased \la\ forest absorption and our normalization
of the continuum.  As we have no knowledge of the continuum shape of
our spectra, we cannot extrapolate a power law slope from the red to
the blue side of \la.  Thus, if absorption increases, our continuum
point on the blue side of the line will be increasingly depressed,
resulting in an over-estimate of the line flux.  The \nv\ line, close
to \la\ will also be effected by this problem, and is generally
dominated by the errors in subtracting the \la\ components.  Previous
work \cite{fk95} has found correlations between the strength of \la\
and redshift, with higher redshift QSOs having weaker \la.  Francis
\& Koratkar attribute this largely to an increase in \la\ forest
absorption in their flux calibrated spectra.  In our composite
spectra, which are continuum divided, this same \la\ forest absorption
causes an underestimate of the continuum level.  This in turn produces
the positive correlation of $\ew$ with redshift.

The \siiv+\oiv] lines had to be fit as one feature, as
they could not be de-blended.  Apart from this they are relatively
clear of contamination.  This blend  shows no evidence whatsoever for
any correlation with luminosity, with equivalent widths which are
constant to $\sim15$\% over a factor of $\sim40$ in luminosity.  This
is in disagreement with Green et al. (2001) who find a significant
correlation with $\ew\propto L^{-0.30\pm0.08}$.  We find a shallower
slope which is consistent with zero.  We note that Green et
al. measure their correlation over a range which is only a factor of
$\sim10$ in luminosity. 

The \civ\ line shows a highly significant correlation of equivalent
width with luminosity, $\ew\propto L^{-0.128\pm0.015}$.  The \ciii\
blend (\ciii, \aliii\ and \siiii) correlation is flatter, with
$\ew\propto L^{-0.070\pm0.008}$. However, this blend also shows
apparent significant correlations with redshift which could be, in
some part, due to the combination of lines taken.  For \mgii, we find
$\ew\propto L^{-0.058\pm0.013}$ with significant departures away from
the correlation at low luminosity, $\mb>-22$.  Checks of the
individual line fits show no obvious systematic problems, and it
appears that these deviations from the simple power law are real.  If
they were also present in the \civ\ or \ciii\ lines and occurred at the
same luminosity we would not see the effect as the faintest QSOs which
have these lines visible are too bright ($\mb=-23$ to $-24$). An
alternative cause is a change in the SED, e.g. of the host galaxy, but
this effect is not seen in the narrow \nev\ emission line which spans
a similar range in luminosity. However, we also note that the
\mgii\ emission line lies on top of the broad \feii\ emission feature
which we have treated as continuum. This \feii\ emission may
contribute a significant proportion ($\sim$25\%) of this continuum
emission and hence any variation in the strength of the \feii\
emission as a function of luminosity will affect the \mgii\ equivalent
width measurements.

Perhaps the biggest surprise of this analysis is the Balmer line
correlation which is markedly different to that of the broad UV
emission lines.  Although the broad \hd\ line shows no significant
correlation, the equivalent widths of the \hg\ and \hb\ lines show a
{\it positive} correlation with luminosity.  There is a hint that the
strength of the correlation increases for lower order Balmer lines
with $\ew\propto L^{0.160\pm0.020}$ and $\ew\propto L^{0.185\pm0.030}$
for \hg\ and \hb\ respectively.  This {\it inverse Baldwin effect} has
not been seen before.  Note that for the \hg\ line we included the
narrow component and the \oiii\ \lam4363 line in the equivalent width,
while for \hb\ we subtract of the narrow component.  However, the
positive correlations with luminosity remain if the narrow \hg\
component is subtracted from the emission line or the narrow \hb\
component is included.

We plot the measured slopes of equivalent width dependence on
luminosity against the ionization potential (filled circles in
Fig. \ref{fig_ionslope}) for all the major emission lines in this
study.  Excluding the \oii\ line, which could be contaminated by
emission from the host galaxy, we see a correlation between the slope
of the Baldwin relation and the ionization potential.

A possible explanation for this effect is that the SED of the ionizing
continuum may steepen towards lower energies with increasing
luminosity, resulting in more photons being available to ionize
hydrogen but relatively fewer with energies greater than 64eV
available to ionize \civ. Alternatively, the ionization parameter may
change as a function of luminosity, as discussed below.
     
On the other hand, the correlations we measure between equivalent
width and luminosity may be caused by changes in the continuum flux under
the lines rather than the line flux itself. We plot the slope
of the equivalent width dependence on luminosity vs. the rest
wavelength of the features in question (Fig. \ref{fig_lamslope}).  In
this case, there may be a trend for the slope of the Baldwin relation
to increase with decreasing wavelength.  However, the broad lines show
a much flatter correlation with wavelength than the narrow lines (even
excluding \oii); arguing against an effect caused by simple continuum 
variation.
 
Our results imply that either (a) L(\hb)/L(\civ) increases
with luminosity, (b) L(4860\AA)/L(1550\AA) decreases with
luminosity or a combination of the two. We have no way of directly
answering this question since our data are not flux calibrated and we
cannot measure the above luminosity ratios. However, we can refer to
earlier findings discussing the line and continuum luminosities in
smaller, less complete samples. 

Earlier studies of AGN SEDs (e.g. Vanden Berk et al. 2001 and
references therein) show that the slope of the power law continuum
changes dramatically near to the \hg\ and \hb\ lines.  Blue-wards of
about 4500\AA\ the continuum slope has $\alpha\sim-0.5$, while
red-wards of this point the slope is more like $\alpha\sim-1.6$. This
has been interpreted as due to the different continuum processes
contributing to the SED at different wavelengths \cite{l90}.  At short
wavelengths most of the emission is due to accretion disks (e.g. Laor
\& Netzer 1989, and references therein) while at longer wavelengths
dust emission, combined with non-thermal emission in radio-loud
sources, is more important. Having this in mind we can speculate that
the relative contribution is luminosity dependent in a sense that the
accretion disk contribution is more important at long wavelengths in
higher luminosity sources. This is exactly the trend observed by Laor
(1990) in his study of the continuum emission in high luminosity AGN.
Other ideas are related to the accretion disk inclination
\cite{n85,nlg92,wilkes99}, changes in the ionizing luminosity as a
function of luminosity \cite{ea99,g98,kbf98,w99} (considered above) or
emission by optically thin gas \cite{sfp95}.  Most of these models
predict a similar trend for all lines albeit with a different
slope. No existing model, except for the one involving changes in the
relative luminosity of the two components, can explain the observed
difference between \civ\ and \hb.

Another possibility is to test the dependence of optical and UV line
ratios, such as L(\hb)/L(\la), on luminosity and continuum shape as
was done for example by Netzer et al. (1995) for a small (20) sample
of radio loud AGN. That study shows a clear correlation of
L(\hb)/L(\la) with L(4861\AA)/L(1216\AA).  However
the small size of the sample, and the small luminosity range, prevents
any clear conclusion.
 
Finally, we note that the \hb\ luminosity range and the \civ\
luminosity range are very different, with almost no overlap, because
of the $z-\mb$ correlation in our sample. It is therefore possible
that the {\it real} Baldwin relationship for {\it all lines} is more
complicated than previously assumed, showing both rising and falling
branches. This may be related to the unusual shape of the $\mb-\ew$
curve of the \mgii\ line seen in Fig. \ref{fig_magbincorr}. This idea
can only be studied by obtaining good \hb\ measurements in high
luminosity AGN.

\section{Conclusions}

We have used composite QSO spectra to make the most accurate
determination of line and continuum correlations to date.  We see the
Baldwin effect  in a number of lines.  In general the equivalent width
correlations are primarily with luminosity and not redshift.  The
broad UV lines generally show strong anti-correlations with
luminosity, although somewhat flatter than previous determinations.
The Balmer line equivalent widths, in contrast, show an inverse
Baldwin effect, and are positively correlated with luminosity.  We
postulate that this difference could be due to a different combination
of disk and non-disk components in AGN of different luminosity.

Some, but not all, narrow forbidden lines also show anti-correlations
with luminosity.  A possible explanation is  that the NLR becomes more
extended, and fainter by comparison, at high luminosity. This has
important implications concerning the possible detection of type 2 QSOs at
high redshifts.  By comparing the strength of the \cak\ absorption
line to the \oii\ emission line via a mean galaxy spectrum we find
that at low luminosities most, if not all, of the \oii\ flux could come
from the host galaxy and not the AGN.  This raises the possibility
that a large fraction of the observed \oii\ in high luminosity AGN is
due to enhanced nuclear star-formation.  Using the \cak\ line and
assuming a constant SED for the host galaxy, we are able to derive the
correlation between host galaxy and AGN luminosity which is
$\lgal\propto\lqso^{0.417\pm0.045}$.

A number of areas have still to be investigated.  First is the
detailed shapes and positions of the lines, which will be discussed in
a forthcoming paper (Corbett et al. in preparation).  Secondly, after
looking at the mean properties as a function of redshift and
luminosity, we should also investigate the variance about this mean by
fitting individual spectra.

\section*{Acknowledgments}

The 2dF QSO Redshift Survey was based on observations made with the
Anglo-Australian Telescope and the UK Schmidt Telescope.  We warmly
thank all the present and former staff of the Anglo-Australian
Observatory for their work in building and operating the 2dF and 6dF
facilities.  KR was supported by an AAO summer vacation studentship
during the course of this work.  HN thanks the director and staff of
the AAO for their hospitality and support during a two month
sabbatical visit in early 2002.

\end{document}